\documentclass[10pt,journal,compsoc]{IEEEtran}
\ifCLASSOPTIONcompsoc
  \usepackage[nocompress]{cite}
\else
  \usepackage{cite}
\fi
\ifCLASSINFOpdf
\else
\fi
\usepackage{booktabs} 

\usepackage{amsmath,amssymb,amsfonts}
\usepackage{algorithmic}
\usepackage{graphicx}
\usepackage{textcomp}
\usepackage{xcolor}
\usepackage{subcaption}
\def\BibTeX{{\rm B\kern-.05em{\sc i\kern-.025em b}\kern-.08em
		T\kern-.1667em\lower.7ex\hbox{E}\kern-.125emX}}

\usepackage{stfloats}
\usepackage{graphicx}
\usepackage[lined,boxed,commentsnumbered, ruled]{algorithm2e}
\usepackage{times,amsmath,amssymb,epsfig,bm}
\usepackage{hyperref}
\usepackage{breakurl}
\usepackage{multirow}
\SetKwComment{Comment}{$\triangleright$\ }{}

\begin{document}
\title{Overcoming Data Sparsity in Group Recommendation}

\author{Hongzhi Yin, Qinyong Wang, Kai Zheng, Zhixu Li, Xiaofang Zhou%
	\IEEEcompsocitemizethanks{\IEEEcompsocthanksitem H, Yin, Q. Wang and X. Zhou are with the School of Information Technology \& Electric Engineering, The University of Queensland, St Lucia, QLD 4072, Australia. E-mail: \{h.yin1, qinyong.wang, zxf\}@uq.edu.au.%
		\IEEEcompsocthanksitem K. Zheng is with the School of Computer Science and Engineering, University of Electronic Science and Technology of China. E-mail: zhengkai@uestc.edu.cn.%
		\IEEEcompsocthanksitem Z. Li is with the School of Computer Science and Technology, Soochow University. E-mail: zhixuli@suda.edu.cn.}
	}

\markboth{IEEE TRANSACTIONS ON KNOWLEDGE AND DATA ENGINEERING, VOL x, NO. x xxxx 2019}%
{Shell \MakeLowercase{\textit{et al.}}: Bare Demo of IEEEtran.cls for Computer Society Journals}
\IEEEtitleabstractindextext{%
\begin{abstract}
It has been an important task  for recommender systems to suggest satisfying activities to a group of users in people's daily social life. The major challenge in this task is how to aggregate personal preferences of group members to infer the decision of a group. Conventional group recommendation methods applied a predefined strategy for preference aggregation. However, these static strategies are too simple to model the real and complex process of group decision-making,  especially for occasional groups which are formed ad-hoc. Moreover, group members should have non-uniform influences or weights in a group, and the weight of a user can be varied in different groups. Therefore, an ideal group recommender system should be able to accurately learn not only users' personal preferences but also the preference aggregation strategy from data.
In this paper, we propose a novel end-to-end group recommender system named \textbf{CAGR} (short for ``\underline{C}entrality-\underline{A}ware \underline{G}roup \underline{R}ecommender''), which takes Bipartite Graph Embedding Model (BGEM),  the self-attention mechanism and Graph Convolutional  Networks (GCNs) as basic building blocks to learn group and user representations in a unified way.
Specifically, we first extend BGEM  to model group-item interactions, and then in order to overcome the limitation and sparsity of the interaction data generated by occasional groups, we propose a self-attentive mechanism to represent groups based on the group members.
In addition, to overcome the sparsity issue of user-item interaction data, we leverage the user social networks to enhance user representation learning, obtaining centrality-aware user representations.
To further alleviate the group data sparsity problem, we propose two model optimization approaches to seamlessly integrate the user representations learning process.  
We create three large-scale benchmark datasets and conduct extensive experiments on them. The experimental results show the superiority of our proposed CAGR by comparing it with state-of-the-art group recommender models.
\end{abstract}

}

\maketitle

\IEEEdisplaynontitleabstractindextext

\IEEEpeerreviewmaketitle

\IEEEraisesectionheading{\section{Introduction}\label{sec:introduction}}

\IEEEPARstart{A}{s}
social animals, people consider group activities as  essential needs for their social life. For example, families often watch TV programs together at night; friends often dine out, watch movies, attend parties and travel together. With the recent development and prevalence of smart phones and social networking services (e.g., Meetup and Facebook Events), it is becoming more convenient and easier for people to get together to form a persistent or occasional group. It is highly urgent to develop group recommender systems to suggest relevant items/events (e.g., dining out, movie watching and parties) for a group of users, known as group recommendation.
The first group recommender system MusicFX~\cite{McCarthy:1998:MAG:289444.289511}  was  developed to recommend music to a group of gym users.
Since then, group recommendation has been seen in various recommendation applications, such as tourism~\cite{mccarthy2006cats} and social events~\cite{Liu:2012:EPI:2396761.2396848}. There are two types of groups: persistent and occasional groups~\cite{Quintarelli:2016:RNI:2959100.2959137,Xiao:2017:FGR:3109859.3109887}. Persistent groups refer to relatively static groups with stable members and sufficient group-item interaction records~\cite{Hu:2014:DMG:2892753.2892811,Said:2011:GRC:2096112.2096113,Ronen:2014:RSM:2600428.2609596,Cao:2018:AGR:3209978.3209998}, such as interest-oriented groups in Meetup; while occasional groups are formed ad-hoc and  users may just constitute the groups for the first time (i.e., cold-start groups)~\cite{Baltrunas:2010:GRR:1864708.1864733,Yuan:2014:CGM:2623330.2623616,Liu:2012:EPI:2396761.2396848}.

To make recommendations to persistent groups, each group can be treated as a virtual user and the personalized recommendation algorithms developed for individual users can be straightforwardly employed since there are sufficient persistent group-item interaction records. However, for occasional groups, their historical interaction data is extremely sparse and even unavailable. Thus, it is infeasible  to directly learn the preference representation of an occasional group, and we can only learn the preferences of an occasional group by aggregating the personal preferences of its  members. In this paper, we focus on a more general scenario of group recommendation, i.e., making recommendations to occasional groups, as the recommendation techniques developed for occasional groups can also be applied to persistent groups.

Group recommendation is much more challenging than  making recommendations to individual users, as different group members may have different preferences. A good group recommendation system should be able to not only accurately learn users' personal preferences, but also model how a decision or consensus  among group members is reached. Prior studies~\cite{Amer-Yahia:2009:GRS:1687627.1687713,Salehi-Abari:2015:PSN:2792838.2800190} on group recommendation systems have been focused on exploring various heuristic aggregation strategies (e.g., average, least misery and maximum pleasure) to find a consensus among group members on an item.  However, all these heuristic and predefined aggregation strategies are too simple to model the real and complex process of group decision-making, leading to suboptimal group recommendation performance. Moreover, a user may exhibit different influences and have different weights in different groups. In this paper, we focus on the essential problem in group recommendation -- preference aggregation, that is how to aggregate personal preferences of group members to decide a group's choice on items.



In our previous work~~\cite{yin2019social}, we proposed a Social Influence-based Group Recommender (SIGR) framework.
In the work, rather than exploring new heuristic and predefined strategies, we introduced the notion of personal social influence to quantify and differentiate the contributions of group members to a group decision, and then proposed to automatically learn the social influence-based aggregation strategy from the group-item  interaction data.
\label{sec:intronew}
However, despite its success, the SIGR model suffers from the following three challenges regarding the model training, model expensiveness, and  sparsity issue of  user-item interaction data. In order to address these three challenges in this paper, we propose a novel solution called \textbf{C}entrality-\textbf{A}ware \textbf{G}roup \textbf{R}ecommender (CAGR) that significantly extends the SIGIR by improving both group and user representation learning. 

\textbf{Challenge 1: End-to-end Learning}. In order to compute the social influence-based weight for each group member,  SIGR first selects and extracts network features from the social network, and then takes these features as input of a deep neural network model to estimate each user's social influence.  This two-stage approach makes it infeasible to train SIGR in an  end-to-end way and leads to suboptimal performance, as each stage has to be optimized separately under different criteria. In addition, it also requires extra supervision.  \textbf{Challenge 2: Model Expressiveness}. SIGR utilizes a plain weighted sum operation to aggregate all raw member representations to obtain the group representation. This linear aggregation function is not expressive enough to capture  the complex interactions and relations among group members and thus cannot accurately mirror the intra-group decision-making process. \textbf{Challenge 3: Data Sparsity of User-Item Interactions}. In order to overcome the data sparsity of group-item interactions, SIGR leverages the user-item interaction data. However, it ignores the sparsity issue of user-item interaction data itself which has been widely recognized in almost all recommendation datasets and also validated in our experiments. In our case, the number of events most users have attended is still far from sufficient to accurately learn user representation or embedding. 

In order to address the above three challenges, we make the following new contributions in our CAGR. 1) Instead of using the social influence-based attention mechanism  developed in~\cite{yin2019social}, we propose to represent the group by simulating the complex interactions among its members and automatically identifying their different group-aware influences based on the self-attention mechanism~\cite{vaswani2017attention}. This method not only significantly enhances the expressiveness of our CAGR, but also facilitates end-to-end model learning that uses a single optimization criterion for enhancing the system and requires  less supervision during training. 2) In order to address the data sparsity issue of user-item interactions, we propose a novel centrality-aware graph convolution module to leverage the social network in terms of homophily and centrality~\cite{zafarani2014social} to enhance user representation learning.
The homophily refers to the tendency  that people are more likely to interact with individuals similar to themselves~\cite{mcpherson2001birds}. 
Due to homophily, in almost all social networks, it has been observed  that neighboring users (e.g., friends) share many common or similar  preferences in respect to a variety of qualities and characteristics.
Meanwhile, the node centrality information is also critical because group decisions are largely influenced or even determined by the group members with high social impacts. The enhanced user representation/embedding can accurately capture both user preference and centrality information, which provides solid foundation for accurately learning group representation.  

To seamlessly integrate the user-item interaction data and  social network data with group-item interaction data, we propose two model optimization approaches to implement our CAGR: a two-stage optimization approach and a joint optimization approach.
Specifically, the two-stage approach first learns embedding of users and items from the user-item interaction data using BGEM in the first stage, which is then utilized to initialize the user/item embedding  in the second stage. We will update the user/item embedding and learn the group embedding  from both  group-item interaction data and social network data in the second stage. The joint approach simultaneously learns the user/item embedding and group embedding from user-item interaction data, group-item interaction data and social network data under the same optimization criteria. The key difference between these two approaches is that there are two objective functions to optimize in the two-stage approach, while there is only one unified objective function to optimize in the joint approach.

The main contributions of this paper are summarized below.

\begin{itemize}
	\item To the best of our knowledge, we are the first work to simultaneously address the critical group-item and user-item data sparsity challenges in the group recommendation task.
	Specifically, we propose a novel self-attention mechanism to aggregate the member embeddings to represent a group.
	Furthermore, we leverage the  social networks to enhance user representation learning through a centrality-aware graph convolution operation.
	\item We propose two model optimization approaches to leverage the user-item interaction data to overcome the limitations and alleviate the sparsity of the group-item interaction data, in which both novel positive sampling approach and negative sampling strategy are developed to advance the conventional stochastic gradient descent algorithm.
	\item We create  three large-scale benchmark datasets for evaluating group recommendation systems, especially the recommenders that are able to make recommendations to occasional groups.  Extensive experiments are conducted to evaluate the performance of our proposed CAGR, and the experimental results show its superiority by comparing with the state-of-the-art techniques.
\end{itemize}

\section{Notations and Problem Formulation}

Technically, our proposed CAGR model consists of two major components: 1) group representation learning that aggregates its member representations with self-attention mechanism on the  group-item data; and 2) centrality-aware user representation learning module that overcomes the user-item data sparsity.  We first present the notations and then formulate the group recommendation problem in this section.

Following the convention, we use bold capital letters (e.g., $\mathbf{X}$) to represent both matrices and graphs, squiggle capital letters (e.g., $\mathcal{X}$) to denote sets, lowercase letters with superscript $\vec{}$  ~(e.g., $\vec{x}$) to denote vectors, normal lowercase letters (e.g., $x$) to denote scalars. All vectors are in column forms if not clarified.  

\begin{figure}[!t]
	\centering
	\includegraphics[width=0.99\columnwidth]{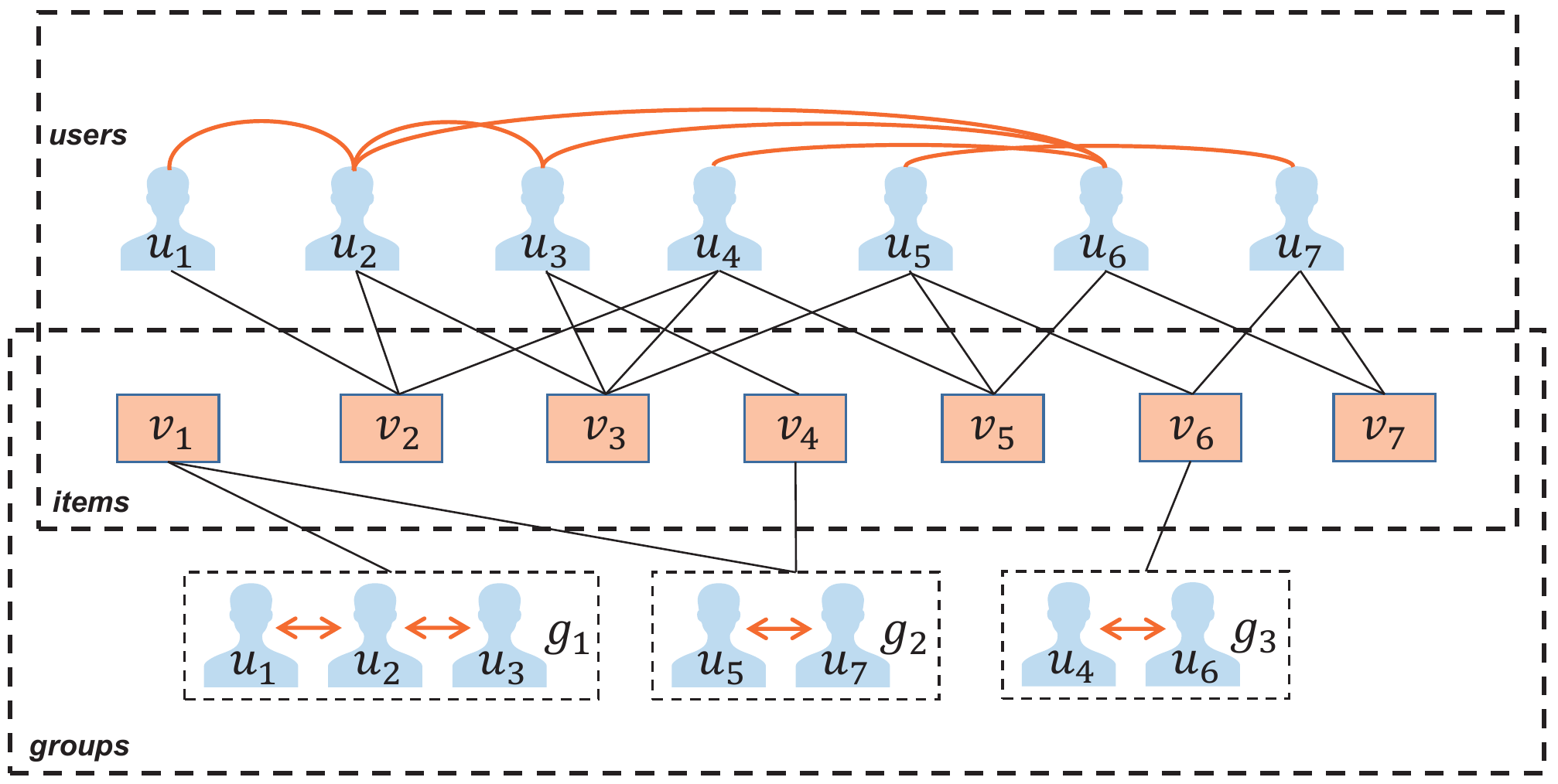}
	\caption{\textbf{Illustration of the input data for the task of making recommendations to occasional groups, including user-item, group-item and user-user interaction data.}}
	\label{fig:task}
	\vspace{-15pt}
\end{figure}

We assume that there are a set of users $\mathcal{U}$, a set of groups $\mathcal{G}$ and a set of items $\mathcal{V}$ in the group recommender system. The $m$-th group $g_{m}\in \mathcal{G}$ consists of a set of users, and we use $\mathcal{G}_{m}$ to denote this set of users. There are three kinds of observed interaction data among  $\mathcal{U}$, $\mathcal{G}$ and $\mathcal{V}$: user-item interactions, group-item interactions and user-user interactions.
We use bipartite graphs $\mathbf{G}_{UV}$ and $\mathbf{G}_{GV}$  to represent user-item interactions and group-item interactions respectively, and use a general graph $\mathbf{G}_{UU}$ to denote user-user interactions, i.e., user social network.
Fig.~\ref{fig:task} illustrates the input data of our group recommendation task.
Then, given an occasional group $g_{m}$, our task is to recommend a ranked list of items that group $g_{m}$ may be interested in, which is formally defined as follows.\\
\textbf{Input}: A set of users $\mathcal{U}$, a set of groups $\mathcal{G}$, a set of items $\mathcal{V}$, group-item interactions $\mathbf{G}_{GV}$, user-item interactions $\mathbf{G}_{UV}$ and user-user interactions $\mathbf{G}_{UU}$.\\
\textbf{Output}: A personalized ranking function that maps an item to a  ranking score for a target occasional group $f_{g}: \mathcal{V}\longrightarrow \mathbb{R}$.


\vspace{-10pt}
\section{Methodologies}
\label{sec:method}

\subsection{Bipartite Graph Embedding Model}

First, we introduce how to extend  BGEM model~\cite{yin2018joint}, which achieves great success  in the bipartite graph embedding problem~\cite{yin2018joint}, to capture the group representations based on group-item interaction data.
The interactions between groups and items can be represented by a bipartite graph $\mathbf{G}_{GV} = (\mathcal{G} \cup \mathcal{V}, \mathcal{E}_{GV})$ where $\mathcal{G}$ is a set of groups and $\mathcal{E}_{GV}$ is a collection of edges between groups and items. If group $g_m$ interacts with item $v_j$, there will be an edge $e_{mj}$ between them. As the rating information of an occasional group is rarely available, we simply set the weight on the edge $e_{mj}$ to be 1.
Given a group $g_m$, we define the probability of $g_m$ interacting with an item $v_j$ as follows:
\begin{equation}
\label{eq:grouppreference}
\small
\centering
p(v_j|g_m)=\frac{\exp(\vec{g}_{m}\cdot \vec{v}_{j})}{\sum_{v_{j'}\in \mathcal{V}}\exp(\vec{g}_{m}\cdot \vec{v}_{j'})},
\end{equation}
where $\vec{g}_{m}$ is the embedding of group $g_{m}$ in the latent space, and $\vec{v}_j$ is the embedding of item $v_j$.
Following the recent word and network embedding techniques~\cite{DBLP:conf/nips/MikolovSCCD13}, BGEM tries to minimize the KL-divergence between the estimated neighbor probability distribution of each group $p(\cdot|g_m)$ and the empirical distribution  $\hat{p}(\cdot|g_m)$. The empirical distribution is defined as $\hat{p}(v_j|g_m)=w_{mj}/d_m$, where $w_{mj}$ is the weight on the edge $e_{mj}$  and $d_m$ is the out-degree of group node $g_m$, i.e., $d_m=\sum_{v_j\in \mathcal{V}}w_{mj}$. By omitting some constants, we obtain the following objective function:
\begin{equation}
\label{eq:groupobjectivefunction}
\small
\centering
O_{GV}=-\sum_{e_{mj}\in \mathcal{E}_{GV}} w_{mj}\log p(v_j|g_m).
\end{equation}
By minimizing the above objective function, we are able to learn each group's embedding $\vec{g}_m$ and each item's embedding $\vec{v}_j$ in a low-dimensional latent space.

To achieve high efficiency in model training, BGEM adopts the approach of negative sampling technique  proposed in~\cite{DBLP:conf/nips/MikolovSCCD13}, which samples multiple negative items to form corrupted examples (or negative edges) according to some noise distribution for each positive example $(g_m,v_j)$. The group learning objective $O_{GV}$ can be reformulated as:
\begin{equation}
\label{eq:groupobjectivefunction3}
\small
\centering
\begin{split}
O_{GV}=&-\sum_{e_{mj}\in \mathcal{E}_{GV}} w_{mj}\big{(}\log \sigma (\vec{g}_{m}\cdot \vec{v}_{j})\\&+ \sum^M_{k=1}E_{v_k\sim P_n(v)}[\log \sigma (-\vec{g}_{m}\cdot \vec{v}_{k})]\big{)},
\end{split}
\end{equation}
where $\sigma(x)=1/(1+\exp{(-x)})$ is the sigmoid function and  $M$ is number of corrupted (negative) examples drawn from a given noise distribution $P_n(v)$.
Following~\cite{DBLP:conf/nips/MikolovSCCD13}, we set the noise distribution $P_n(v)\propto d_{v}^{0.75}$, where $d_{v}$ is the out-degree of item node $v$. 

\subsection{Self-Attentive Group Representation Learning}
\label{st:att}
However, it is infeasible to directly learn the embedding of an occasional group $\vec{g}_{m}$  from the group-item interaction data due to the cold-start nature of occasional groups. In contrast to persistent groups,  an occasional group is defined as a number of persons who do something occasionally together, like having a dinner, watching a movie, attending a party and visiting a POI~\cite{boratto2010state}. Its members have a common aim only in a particular moment. There are many contexts where  a group of persons is not established for some shared long-term interests, but might be occasionally interested in getting together for a common aim, e.g., people attending events together or traveling together. As occasional groups are typically short-lived by definition and many new occasional groups are being created, they often have little or no historical interaction data.  The  group-item interaction matrix is much sparser than user-item matrix (referring to relevant statistics of three real-life datasets in Table I). The problem of cold-start groups  arises naturally, and the classic group recommendation techniques~\cite{Hu:2014:DMG:2892753.2892811,Cao:2018:AGR:3209978.3209998} that assume groups have ample historical interaction records would significantly underperform in this scenario.

To address the cold-start problem, we propose to learn a group's embedding  by aggregating the embedding of its members. Specifically, given an occasional group $g_m$, its embedding is represented as follows:
\begin{equation}
\label{eq:groupaggfa}
\centering
\vec{g}_m=f_{a}(\vec{u}_{i} | \forall u_i \in \mathcal{G}_m),
\end{equation}
where $f_{a}(.)$ is the aggregating function, $\mathcal{G}_m$ represents the set of users who constitute  group $g_m$ and $\vec{u}_{i}$ is the embedding of group member $u_i$. 
As group members have different social statuses, expertise, reputation, personality and other social factors~\cite{7436655,alina2014social}, they are not equal and have different social influences in the group's decisions and choices.
A user can exhibit different social influences in different groups that consist of different members. An important aspect of group activities is the need to reach consensus. In non-virtual environments, consensus results from negotiation among group members, especially those group members with low  social influences are often willing to modify their initial individual opinions and compromise to satisfy the preferences of the influential members. Sometimes, a group's preferences  reflect the preferences of a few influential members (e.g., group leaders or opinion leaders) rather than the common preferences of most group members.

Motivated by this, we proposed a social influence-based group representation framework SIGR in our previous work~\cite{yin2019social} that adopts the linear weighted sum over the embeddings of its members as the aggregating function $f_a(.)$: 
\begin{equation}
\label{eq:groupagg}
\centering
\vec{g}_m=f_{a}(\vec{u}_{i} | \forall u_i \in \mathcal{G}_m) = \sum_{u_i\in \mathcal{G}_m}\lambda_{im}\vec{u}_{i},
\end{equation}
where the attention weight $\lambda_{im}$  denotes $u_i$'s social influence/weight in group $g_m$, and it also reflects how much $u_i$ contributes to the group's decision-making.
The challenge is how we can learn the attention weight $\lambda_{im}$.
As occasional groups have few  historical interactions on items, it is infeasible to directly learn the group-aware personal social influence $\lambda_{im}$. 
In~\cite{yin2019social}, we introduced a non-negative latent variable to represent the global social influence of user $u_i$, which is independent from specific groups and used to compute $\lambda_{im}$ based on the vanilla attention mechanism~\cite{bahdanau2014neural}. 
In order to learn this global latent variable,
we proposed a two-stage approach in~\cite{yin2019social}. We first precomputed both global  centrality features (e.g., PageRank centrality, closeness centrality, betweenness centrality and eigenvector centrality) and local structure features of each node, and then took these network features as the input of a neural network model to estimate the global latent variable value for each user node. Note that we adopted network embedding models such as DeepWalk~\cite{Perozzi:2014:DOL:2623330.2623732} and node2vec~\cite{Grover:2016:NSF:2939672.2939754} to extract local structure features in the form of node embedding.



However, the social influence-based attention mechanism faces two major challenges. First, it is not an end-to-end solution and requires extra supervision during training, which has the risk of damaging the overall system performance. Second, it is not expressive enough to accurately model the complex intra-group interactions to infer final group decisions because it uses a plain linear weighted sum aggregation function over the raw group member embeddings.

Therefore, we propose a method in which we represent $\vec{g}_m$ in Equation~\ref{eq:groupaggfa} by purely modeling the complicated intra-relations among its members based on the self-attention mechanism~\cite{vaswani2017attention,parikh2016decomposable}, which can effectively relate different positions in a single list  to compute its representation. Another benefit of this method is that it is end-to-end because we do not need to extract social network features  in advance to calculate the attention weights, which could avoid extra supervision required by the two-stage method in SIGR and enhance the system. 

Given a $d$-dimensional user representation $\vec{u}_{i}$, we project it into  three different $d$-dimensional vectors:
the query vector, the key vector  and the value vector, by multiplying $\vec{u}_{i}$ with three corresponding trainable   embedding matrices.
To benefit from the efficiency of matrix computation, we pack together all query, key and value vectors transformed from the users in the same group (i.e., $\{\vec{u}_{i} | \forall u_i \in \mathcal{G}_m\}$) into three matrices $\mathbf{Q} \in \mathbb{R}^{ |\mathcal{G}_m| \times d}$, $\mathbf{K} \in \mathbb{R}^{ |\mathcal{G}_m| \times d}$ and $\mathbf{V} \in \mathbb{R}^{ |\mathcal{G}_m| \times d}$.
With $\mathbf{Q}$, $\mathbf{K}$ and $\mathbf{V}$, we obtain the output matrix $\mathbf{O}  \in \mathbb{R}^{ |\mathcal{G}_m| \times d}$, which is a weighted sum of the values, where the weight assigned to each value is determined by the dot-product of the query with all the keys:
\begin{equation}
\label{eq:singlehead}
\begin{split}
\mathbf{O} &= \textrm{Attention}(\mathbf{Q}, \mathbf{K}, \mathbf{V}) \\
&= softmax(\frac{\mathbf{Q\mathbf{K}^T}}{\sqrt{d}} ) \mathbf{V},
\end{split}
\end{equation}
where $\mathbf{O}$ consists of the transformed user embeddings which additionally capture the information of interactions and relations among the group members.

Noting that the above self-attention mechanism only performs one single attention function to the queries, keys and values, we further improve its efficiency and expressiveness by adopting the multi-headed attention method~\cite{vaswani2017attention}.
This extension is quite important because it allows the model to jointly attend to information from different representation
subspaces at different positions, and thus enhance its expressiveness power.
Another potential benefit is that these projections can be efficiently computed in parallel~\cite{vaswani2017attention}.

The multi-head attention mechanism obtains $h$ (i.e. one per head) different representations of $(\mathbf{Q}, \mathbf{K}, \mathbf{V})$, computes scaled dot-product attention for each representation, concatenates the results, and projects the concatenation through a feed-forward layer.
Specifically, for the $i$-th head $\mathbf{M}_i$, we follow the aforementioned approach to construct and transform the queries, keys and values (i.e., $\mathbf{Q}$, $\mathbf{K}$ and $\mathbf{V}$) by three trainable matrices $\mathbf{W}_i^Q \in \mathbb{R}^{d \times \frac{d}{h}}$, $\mathbf{W}_i^K \in \mathbb{R}^{d \times \frac{d}{h}}$ and $\mathbf{W}_i^V \in \mathbb{R}^{d \times \frac{d}{h}}$, respectively:
\begin{equation}
\label{eq:multitrans}
\mathbf{M}_i^Q = \mathbf{Q}\mathbf{W}_i^Q; \quad
\mathbf{M}_i^K = \mathbf{K}\mathbf{W}_i^K; \quad
\mathbf{M}_i^V = \mathbf{V}\mathbf{W}_i^V. 
\end{equation}
Then the  self-attention mechanism is used to compute the relevance between queries and keys, and output the mixed representations.
\begin{equation}
\mathbf{M}_i = \textrm{Attention}(\mathbf{M}_i^Q, \mathbf{M}_i^K, \mathbf{M}_i^V).
\end{equation}
After that, we concatenates all those heads, and fed it through a feed-forward layer.
\begin{equation}
\label{eq:multihead}
\mathbf{O}  = \textrm{Concat}(\mathbf{M}_1, \ldots, \mathbf{M}_h)\mathbf{W}^O,
\end{equation} 
where $\mathbf{W}^O \in \mathbb{R}^{d \times d}$ is a trainable matrix.

Finally, inspired by~\cite{yang2016hierarchical}, we again apply the vanilla attention mechanism to the output matrix $\mathbf{O}$ to select the influential users and form the group representation $\vec{g}_m$:
\begin{equation}
\label{eq:attentgroup}
\begin{aligned}
\vec{a}_{i} &=\textrm{tanh} (\mathbf{W_s} \vec{O}_{i}+\vec{b}_{s}), \\
\lambda_{i} &=\frac{\exp (\vec{a}_{i}^{\top} \vec{a}_{s})}{\sum_{j} \exp (\vec{a}_{j}^{\top} \vec{a}_{s})}, \\
\vec{g}_m &= f'_{a}(\vec{u}_{i} | \forall u_i \in \mathcal{G}_m) = \sum_{i} \lambda_{i} \vec{a}_{i},
\end{aligned}
\end{equation}
where $\mathbf{W_s}$ and $\vec{b}_{s}$ are the parameters for the feed-forward network, $\textrm{tanh}(.)$ is the hyperbolic tangent activation function, and $\vec{a}_{s}$ is the group level context vector and can be randomly initialized and jointly learned during the training process.

By introducing the  aggregation function to our group representation learning, we update the basic objective function $O_{GV}$ to $O_{SGV}$ as follows:
\begin{equation}
\label{eq:groupobjectivefunction4}
\small
\centering
\begin{split}
O_{SGV}&=-\sum_{e_{mj}\in \mathcal{E}_{GV}} w_{mj}\big{(}\log \sigma (f'_{a}(\vec{u}_{i} | \forall u_i \in \mathcal{G}_m))\cdot \vec{v}_{j})\\&+ \sum^M_{k=1}E_{v_k\sim P_n(v)}[\log \sigma (-f'_{a}(\vec{u}_{i} | \forall u_i \in \mathcal{G}_m))\cdot \vec{v}_{k})]\big{)}.
\end{split}
\end{equation}

\vspace{-8pt}
\subsection{Centrality-Aware User Representation Learning}
\label{st:up}

Directly optimizing the above objective function $O_{SGV}$ in Equation~\ref{eq:groupobjectivefunction4} would lead to inaccurate user and group embeddings due to the sparsity of group-item interaction data. 
Therefore, we propose to enhance the representation learning of users from other auxiliary data sources, which further improves the quality of  group representation learning.

To alleviate the group data sparsity issue, the SIGR model~\cite{yin2019social}  simply applies the BGEM model~\cite{yin2018joint} to the user-item interaction data to enhance user representation learning.
However, this approach ignores the data sparsity issue of user-item interactions that has been widely recognized by the community of recommender systems and also evidenced by the relevant statistics in  Table~\ref{tb:stats}, even though an
individual user usually interacts with more items compared to a
group.
To address this challenge, we exploit and integrate  the social network to further enhance user representation learning motivated by the phenomenon of assortativity in social networks~\cite{newman2002assortative}.
On the other hand, the social networks provide strong signals about users' global social influences~\cite{Liu:2012:EPI:2396761.2396848,7436655}, which are critical factors for modeling the decision-making process within groups as discussed in~\cite{yin2019social}. 
However, current network embedding techniques including BGEM focus on  capturing only the low or high-order proximity between vertices, and ignore the crucial centrality information that indicates
how important/influential a user is in the social network. 
Therefore, it is necessary to encode both global node centrality and local neighborhood information in the social networks into the user representations via an end-to-end and computationally efficient manner.
In what follows, we give a detailed description on how to enhance the user representations 
by exploiting and integrating the centrality information in the social network based on the GraphCSC~\cite{chen2019exploiting} model.

Given a  user-user social network $\mathbf{G}_{UU} = (\mathcal{U}, \mathcal{E})$ where $\mathcal{U}$ and $\mathcal{E}$ are the sets of users and their social ties.
Following GraphCSC, we define a  graph convolution operation, whose core idea is to aggregate information from the neighborhood of a given node $i$.
This procedure is listed in Algorithm 1 CONVOLVE.
Specifically, in Line 1, we transform the representations of user $i$'s neighbors $\{\vec{u}_n | \forall n \in \mathcal{N}_i\}$ through a dense neural network parameterized by the  weights matrix $\mathbf{P}$ and bias vector  $\vec{p}$, and activated by the rectified linear unit $\textrm{Relu}$~\cite{nair2010rectified}.
The Relu activation function is employed due to its two advantages over the sigmoid/tanh functions. First, it enables faster convergence in the SGD optimization process~\cite{krizhevsky2012imagenet}. Second, it does not involve expensive operation because it can be implemented by simply thresholding a matrix of activations at zero.
Then we apply an aggregator/pooling function such as an element-wise
mean or weighted sum on the resulting set of vectors to 
obtain a vector representation $\vec{h}_i$ for the local neighborhood.
In Line 2, we concatenate the aggregated neighborhood vector $\vec{h}_i$ with user $i$'s current representation $\vec{u}_i$ and transform the concatenated vector through another dense neural network layer parameterized by the  weights matrix $\mathbf{W}$ and bias vector  $\vec{w}$, and also activated by $\textrm{Relu}$,
where the concatenation operation instead of the average operation is employed due to its significant improvement shown in~\cite{kipf2016semi}.
Finally, Line 3 normalizes the representations.

\begin{figure}[!t]
	\centering
	\includegraphics[width=\columnwidth]{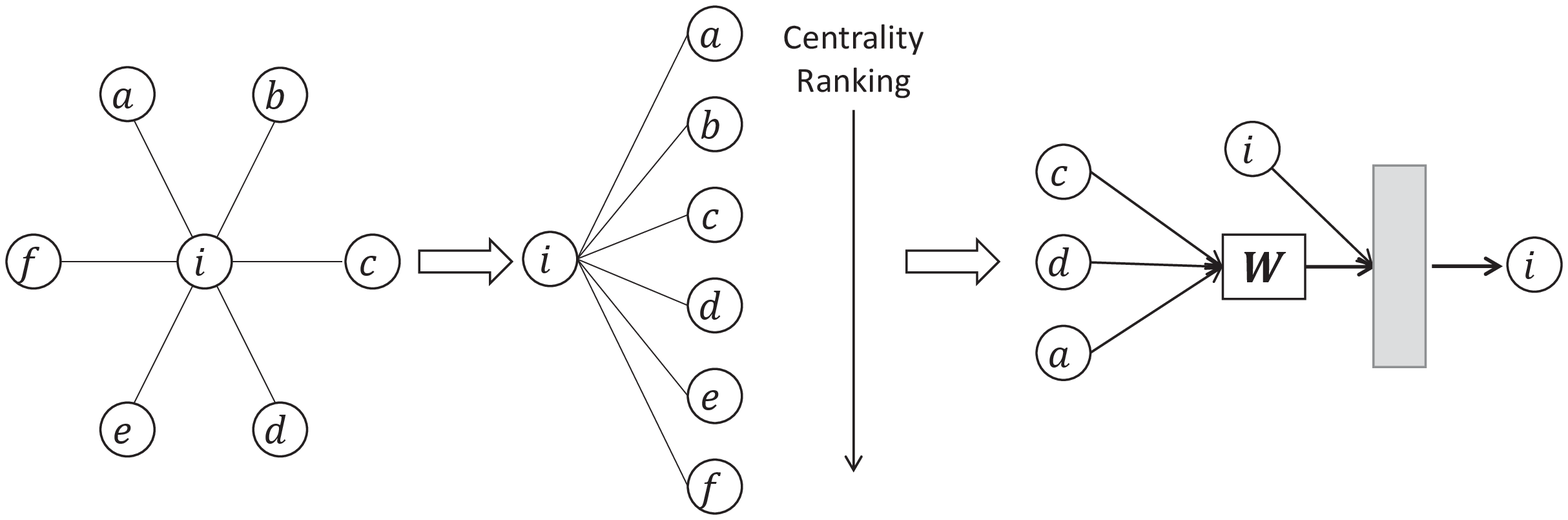}
	\caption{\textbf{The simplified process to generate a centrality-aware network embedding based on the algorithm CONVOLVE.}}
	\label{fig:gcn}
	\vspace{-15pt}
\end{figure}

The output of the algorithm is a representation of node $i$ that incorporates both information about itself and its local graph
neighborhood.
How to select the   neighbors to enhance the user representation learning is the key to the performance of this algorithm. 
In order to incorporate the centrality information, we sample the neighbors of a node based on their scores of a given  centrality measurement $c$.
Specifically, we first rank all the neighboring nodes w.r.t. $c$, then we only consider a fixed number of the top ranked nodes to perform the convolution operation.
In this way, the centrality information measured by $c$ is encoded into the eventual node representations. A simplified illustration of this process is shown in Fig.~\ref{fig:gcn}.

\begin{algorithm}[!t] \label{algo:conv}
	\caption{CONVOLVE} \BlankLine
	\LinesNumbered
	\KwIn{Current embedding $\vec{u}_i$ for user $i$, embeddings of sampled neighbors $\{\vec{u}_n | n \in \mathcal{N}_i$\};} \BlankLine
	\KwOut{New embedding $\vec{u}^{NEW}_i$ for user $i$;}  \BlankLine
	
	$\vec{h}_i \gets POOLING(\{\textrm{Relu}(\mathbf{P} h_{n}  + \vec{p}) | \forall n \in \mathcal{N}_i\})$ ;\\
	\BlankLine
	$\vec{u}^{NEW}_i \gets \textrm{Relu}(\mathbf{W}\cdot \textrm{Concat}(\vec{u}_i, \vec{h}_i)  + \vec{w})$; \\
	\BlankLine
	$\vec{u}^{NEW}_i \gets \vec{u}^{NEW}_i /  \| \vec{u}^{NEW}_i \|_2 $
\end{algorithm}

Moreover, we know that different centrality measurements play different roles when describing a node from different perspectives. For example,  betweenness centrality~\cite{zafarani2014social} counts the fraction of shortest paths going through a node and PageRank centrality~\cite{page1999pagerank}
works by counting the number and quality of links to a node to determine a rough estimation of how important the node is. Therefore, it is beneficial to include multiple centrality information, rather than a single one, to the embeddings, and we employ the multi-view method to achieve this.
Given multiple centrality measurements $\mathcal{C}=\{c_1, c_2, \ldots\}$, we define  centrality oriented views for the graph $\mathbf{G}_{UU}$ by ordering the network vertices based on each of these centrality measurements, which  enables corresponding view-specific graph convolutions. Each of these convolutions follows the aforementioned method to generate the node embeddings but with different centrality rankings to form the receptive field.
We denote the user representation for user $u_i$ under the centrality-specific view associated with the measurement $c_k$ as $\vec{u}_i^{c_k}$. 
The final robust user representation is calculated as the weighted combinations of
the centrality-specific representations with coefficients as the voting weights of centrality-specific views:
\begin{equation}
\label{eq:userrepresentation}
\vec{u}_i=\sum_{k=1}^{|\mathcal{C}|} \alpha_{i}^{c_k} \vec{u}_i^{c_k},
\end{equation}
where $\alpha_{i}^{c_k}$ is the weight for $\vec{u}_i^{c_k}$ in view $c_k$, and is calculated via a Softmax function:
\begin{equation}
\alpha_{i}^{c_k}=\frac{\exp \left(\vec{z}_{k} \cdot \vec{u}_{i}^{C}\right)}{\sum_{j} \exp \left(\vec{z}_{j} \cdot \vec{u}_{i}^{C}\right)},
\end{equation}
where  $\vec{u}_{i}^{C}$ is the concatenation of all centrality-specific representations
of node $i$, and $\vec{z}_{k}$ is a trainable feature vector of centrality-specific
view $c_k$, describing what centrality information has a strong impact on certain nodes. If the dot product between feature vector $\vec{z}_{k}$ and concatenated representations $\vec{u}_{i}^{C}$ is large, it means that centrality $c_k$ is informative for $u_i$, correspondingly, the weight of this view is relative large.

After the convolution operation is applied to all user nodes, we can obtain their centrality-aware representations.
These user representations are learned on the user-item interaction data following the BGEM framework to get the final user embeddings.
We denote this learning objective function as $O_{UV}$:
\begin{equation}
\label{eq:objectivefunction3}
\small
\centering
\begin{split}
O_{UV} & =-\sum_{e_{ij}\in \mathcal{E}_{UV}} w_{ij}\big{(}\log \sigma (\sum_{k=1}^{|\mathcal{C}|} \alpha_{i}^{c_k} \vec{u}_i^{c_k}\cdot \vec{v}_{j})\\ &+ \sum^M_{k=1}E_{v_k\sim P_n(v)}[\log \sigma (-\sum_{k=1}^{|\mathcal{C}|} \alpha_{i}^{c_k} \vec{u}_i^{c_k}\cdot \vec{v}_{k})].
\end{split}
\end{equation}


\begin{algorithm}[!t] \label{algo:train}
	\small
	\caption{Joint Training Algorithm} \BlankLine
	\LinesNumbered
	\KwIn{$\mathbf{G}_{UV}$, $\mathbf{G}_{GV}$, $\mathbf{G}_{UU}$, number of iterations $N$, number of negative samples per positive sample $M$, the sampler parameter $\eta$;} \BlankLine
	\KwOut{The well-tuned model parameter set} \BlankLine
	$iter\leftarrow 0$;\\
	\While{$iter$ $\le$ $N$}
	{
		Flip a coin $c$ according to a $bernoulli(\frac{1}{1+\eta})$;\\
		\If {$c=1$}
		{
			Randomly draw a positive edge $e_{ij}\in \mathcal{E}_{GV}$;\\
			Sample $M$ negative edges for $e_{ij}$;\\
			Update the associated model parameters based on the gradients w.r.t. Equation~\ref{eq:groupobjectivefunction4}  ;\\
		}
		\Else
		{
			Randomly draw a positive edge $e_{ij}\in \mathcal{E}_{UV}$;\\
			Sample $M$ negative edges for $e_{ij}$;\\
			Get the representation of related user nodes from the social network based on Equation~\ref{eq:userrepresentation};\\
			Update the associated model parameters based on the gradients w.r.t. Equation~\ref{eq:objectivefunction3};\\
			
		}
		$iter$=$iter$+1;\\
	}
\end{algorithm}

\section{Model Optimization}
\label{sec:opti}

By optimizing the group learning objective function $O_{SGV}$ in Equation~\ref{eq:groupobjectivefunction4}, we are only able to obtain the embedding of users who have attended at least one group activity and embedding of items that interact with at least one group.
However, due to the cold-start nature of occasional groups, they may contain members who have never attended any group activity before. Besides, the group-item interaction matrix is very sparse, thus the embedding of both users and items and  parameters of the aggregating function learned by optimizing  $O_{SGV}$  is not accurate or reliable.  To effectively overcome the limitations and sparsity issue of the group-item interaction data, we propose to additionally learn the user embeddings from the user activity data, i.e., the user-item and user-user interaction data.
Technically, we develop two model optimization approaches to integrate $O_{UV}$ with $O_{SGV}$: Two-stage Training and Joint Training.

\textbf{Two-stage Training}. We first optimize the objective function $O_{UV}$ to obtain the embedding of users and items ($\vec{u}_{i}$ and $\vec{v}_{j}$) in the first stage.  In the second stage, these learned embeddings are taken as the initial values of the embedding of users and items in $O_{SGV}$, and then they will be also fine-tuned and updated during the process of optimizing the objective $O_{SGV}$. Specifically, we adopt the Stochastic Gradient Descent algorithm (SGD) algorithm to optimize $O_{SGV}$. In each gradient step, we randomly sample a positive example $(g_m, v_j)$ and $M$ negative examples $(g_m,v_k)$ to update model parameters.

\textbf{Joint Training}. By combining objectives $O_{UV}$ and $O_{SGV}$, the joint objective function can be simply defined as follows:
\begin{equation}
\label{eq:jointobjective}
\small
\centering
O_{GUV}=O_{SGV}+ O_{UV}.
\end{equation}
To optimize the above joint objective function, we cannot straightforwardly use SGD, because $O_{SGV}$ and $O_{UV}$ in Equation~\ref{eq:jointobjective} have different training instances: group-item pairs vs. user-item pairs. To address this issue, one possible solution is to first merge all edges in  edge sets  $\mathcal{E}_{UV}$ and $\mathcal{E}_{GV}$ into a big edge set, and then randomly sample a positive edge from the merged edge set in each gradient step, just as done in~\cite{Xiemincikm:2016,yin2018joint}.
However, the group-item interaction graph is much sparser than the user-item interaction graph, i.e., the number of edges in $\mathcal{E}_{GV}$ is much smaller than the  number of edges in $\mathcal{E}_{UV}$. If we uniformly draw a positive edge from the merged edge set to perform stochastic gradient descent, most of sampled positive edges would be user-item edges, and there would not be enough group-item interaction edges for accurately estimating the trainable parameters of the aggregating function.
To overcome the challenge of data skewness, we propose a novel joint training procedure in Algorithm~2. Instead of merging all edges  into a big edge set, we will first draw or choose a bipartite graph with the sampling probabilities $\frac{1}{1+\eta}$ and  $\frac{\eta}{1+\eta}$  for the group-item graph and  user-item  graph respectively,  and then randomly draw a positive edge and $M$ negative edges from the sampled bipartite graph to update the gradients. By doing so, the joint objective function is actually changed to the following equation:
\begin{equation}
\label{eq:improvedjointobjective}
\small
\centering
O_{GUV}=O_{SGV}+ \eta O_{UV},
\end{equation}
where $\eta$ is a non-negative hyper-parameter that is used to control the weight or contribution of the objective  $O_{UV}$.

\textbf{Time Complexity Analysis}.  For each stochastic gradient step in Algorithm~2, the time complexity for the convolution operation  is small and can be ignorable~\cite{chen2019exploiting}, so the time complexity   is $O(d\cdot M\cdot |\mathcal{C}|)=O(d)$, where $M$ and $|\mathcal{C}|$ are often small (less than 10) in large-scale datasets~\cite{yin2018joint} and thus can also be ignorable;  $d$ is the embedding dimension  and also typically small. We assume that our model needs $N$ samples (i.e., $N$ stochastic gradient steps) to reach convergence, thus its overall time complexity is $O(d\cdot N)$.
In practice, the required  number of stochastic gradient steps $N$  is typically proportional to the number of edges~\cite{yin2018joint}.

\vspace{-5pt}
\subsection{Negative Sampling of Items}

How to sample $M$ negative items to form $M$ negative edges (i.e., corrupted examples) for each positive edge (i.e., each observed edge)?  For a positive user-item edge $(u_i,v_j)$ on $\mathbf{G}_{UV}$, we employ the widely adopted degree-based noise distribution $P_n(v_k)\propto d_{v_k}^{0.75}$~\cite{DBLP:conf/nips/MikolovSCCD13}, where $d_{v_k}$ is the out-degree of item node $v_k$ on the user-item graph.
However, this classic negative sampling method does not apply to the occasional group-item interaction graph $\mathbf{G}_{GV}$, because the group-item graph is extremely sparse and the variance of its node degrees is not so obvious. In this case, the degree-based negative sampling strategy  may degrade to the uniform negative sampling.
Most negative examples generated in this way are ``too easy'' and will contribute little to learning an effective discriminator, because they are  obviously false.
To generate more difficult and informative negative examples for each positive edge $(g_m,v_j)$ on $\mathbf{G}_{GV}$, we propose a novel group-aware negative sampling technique by leveraging the user-item interaction graph. Specifically, given a group $g_m$,  the noise distribution is changed to be group-aware, i.e., $P^{g_m}_n(v_k)\propto (d^{g_m}_{v_k}+\gamma)^{0.75}$, where $d^{g_m}_{v_k}$ represents the popularity of item $v_k$ among its members, which can be easily derived from the user-item interaction graph $\mathbf{G}_{UV}$ as follows $d^{g_m}_{v_k}=\sum_{u_j\in \mathcal{G}_m} w_{jk}$; $\gamma$ is a smoothing constant parameter which assigns a small probability to items that have no interactions  with its group members. Thus, the generated negative items in this way will be more popular among the group's members,  and they are   more informative and helpful  to learn the discriminative weights in the self-attention mechanism.

%
\subsection{Group Recommendation using CAGR}

Once we have learned the model parameters in CAGR, given an occasional group  $g_m$, we first use the aggregation function defined in Equation~\ref{eq:attentgroup} to aggregate its members (i.e. $u_i\in \mathcal{G}_m$), obtaining the group representation $\vec{g}_{m}$. Then, a ranking score for each item $v_j$ can be computed according to the dot product of $\vec{g}_{m}$ and $\vec{v}_j$, i.e., $s_{g}(g_m,v_j)=\vec{g}_{m}\cdot\vec{v}_j$.
Finally the top-$n$ items with highest ranking scores will be recommended to the group $g_m$.

\vspace{-5pt}
\section{Experiment Setup}

In this section, we introduce the experimental settings, including research questions to answer, datasets, evaluation protocols and comparison methods.
\vspace{-5pt}
\subsection{Research Questions}

We conduct extensive experiments on three large-scale benchmark datasets to answer the following research questions and validate our technical contributions.\\
\textbf{RQ1:} How does our proposed group recommender model CAGR perform  compared with state-of-the-art group recommenders and various predefined aggregation strategies? \\
\textbf{RQ2:}  Can we improve the group recommendation  by using GCNs-based techniques which leverage the social network structure information to enhance user embedding learning? If the answer is yes, can the centrality-aware GCNs method GraphCSC achieve better performance than the vanilla GCNs method~\cite{kipf2016semi} in capturing user influence from the social network?\\
\textbf{RQ3:} Can we improve the group recommendation  by integrating the user activity data, i.e., the user-item interaction and user social network data? If yes, how do our proposed two model optimization approaches perform on heterogeneous interaction data? Furthermore, for the joint optimization approach, which negative sampling strategy is more suitable for the group-item interaction data? \\
\textbf{RQ4:} How do some of the most important hyper-parameters (e.g., $d$ and $M$) affect the performance of CAGR?

Besides those four research questions, we are also interested in the capability of BGEM in modeling and predicting interactions on the task of top-$n$ recommendation for individual users (i.e., user-item interaction prediction). We have already conducted this experiment and showed the results in our previous work~\cite{yin2019social}.
In the study, we compared BGEM with some strong baseline models such as \textbf{BPR}~\cite{rendle2009bpr}, \textbf{eALS}~\cite{he2016fast} and \textbf{NCF}~\cite{he2017neural} on the user-item interaction data for making recommendations to individual users, and the experimental results showed that BGEM outperforms those baseline methods. 
This justifies the choice of BGEM as a fundamental building block of our proposed SIGR and CAGR models. We refer the readers interested in this study to~\cite{yin2019social} for more details.


%
%
%

\vspace{-5pt}
\subsection{Datasets}

\begin{table}[!t]
	\small
	\caption{Basic Statistics of the Three Datasets.}
	\centering
	\begin{tabular}{cccc}
		& Yelp & Douban-Event & Meetup 
		\\\hline
		\footnotesize{\# Users} &34,504 &70,743 & 24,631\\
		\footnotesize{\# Groups} & 24,103 &110,597 &13,552\\
		\footnotesize{\# Items} &22,611 &60,028 &19,031\\\hline
		\footnotesize{Avg. group size} &4.45   & 4.82 &8.79\\
		\footnotesize{Avg. \#interactions for a group} &1.12  & 1.48 &1.40\\
		\footnotesize{Avg. \#interactions for a user} &13.98 & 48.38 &5.15\\
		\footnotesize{Avg. \#friends for a user} &20.77 & 86.08 &35.75\\
		\hline
	\end{tabular}
	\label{tb:stats}
\end{table}

As existing publicly available group recommendation datasets such as CAMRa2011\footnote{\url{http://2011.camrachallenge.com/2011}} and Movielens-Group~\cite{Yuan:2014:CGM:2623330.2623616} consist of either a smaller number of persistent groups or randomly generated groups and they do not contain the user social network information, they are not suitable to evaluate our solution  CAGR. This is why we need to create three large-scale benchmark group recommendation datasets based on the Yelp Challenge dataset\footnote{\url{https://www.yelp.com/dataset/challenge}}, Douban-Event dataset~\cite{yin2018joint} and Meetup\footnote{\url{https://www.kaggle.com/stkbailey/nashville-meetup}}.
Douban Event is the largest online event-based social network in China that helps people publish and participate in social events. For each user,
we acquired her event attendance list and social friend list. For each event, its time and venue were also collected.
Yelp allows users to share their check-ins about local businesses (e.g., restaurants) and create social connections with other users.  Each check-in or review contains a user, a timestamp and a business, indicating the user visited the business at that time.  In our Yelp dataset, we only focus on the restaurants located in the Los Angeles area, where there are 34,504 users and 22,611 restaurants.
Meetup is an online social event service where users can publish and participate in social events. On Meetup, a social event is created by a user who specifies the time, location and event description.
Other users may express their interests of attending this event.

As Yelp and Douban-Event do not contain explicit  group information,  we extract implicit group activities
as follows: we assume if a set of users who are connected on the social network visit the same restaurant  at the
same time or attend the same event, they are the members of a group and the corresponding activities are group activities.
Meetup contains explicit group information, but it does not have explicit social network data. So we follow the approach proposed in~\cite{liu2012event} to form the social network.
The statistical information of the three datasets is shown in Table~\ref{tb:stats}. From the table, we can see that group-item interaction data is much sparser than the user-item interaction data. For example, in the Douban-Event dataset, a group has only 1.48 interaction records on average, but a user has 48.38 interaction records.

All datasets used in our work are publicly available\footnote{\url{https://sites.google.com/site/dbhongzhi}}.

\vspace{-5pt}
\subsection{Evaluation Methodology}
To evaluate the performance of  group recommendation systems, we first rank all group-item interaction records according to their timestamps in each dataset, and then
use the 80-\textit{th} percentile as the cut-off point so that the group-item interactions before this point will be used for training, and the rest are for testing. In the training dataset, we choose the last 10\% records as the validation data to tune the model hyper-parameters such as $\eta$ and $M$.
According to the above dividing strategies, we split the group-item interaction records in each dataset $\mathcal{D}$ into the training set $\mathcal{D}_{training}$ and the test set $\mathcal{D}_{test}$.

We employ the widely adopted metric  \emph{Hits ratio}~\cite{cremonesi2010performance,hu2013spatial,yin2018joint,Wang2018} to measure the recommendation accuracy. Specifically, for each group-item interaction $(g, v)$ in the test set $\mathcal{D}_{test}$:\\
(1) We compute a ranking score for item $v$ as well as other items that group $g$ has never interacted with.\\
(2) We form a top-$n$ recommendation list by picking $n$ items with the highest ranking scores. If the ground-truth item $v$ appears in the top-$n$ recommendation list, we have a \emph{hit}. Otherwise, we have a \emph{miss}.

The metric Hits ratio is defined as follows:\\
\begin{equation}
\label{eq:hits}
\small
\centering
Hits@n=\frac{\#hit@n}{|\mathcal{D}_{test}|},
\end{equation}
where $\#hit@n$ denotes the number of \emph{hits} in the test set, and $|\mathcal{D}_{test}|$ is
the total number of test cases in the test set. A good group recommender model should achieve higher Hits$@n$.

Besides Hits ratio, we also adopt the commonly used metric Mean Reciprocal Rank (MRR) to measure the recommendation accuracy, and it is defined as follows:
\begin{equation}
\label{eq:hits}
\small
\centering
MRR=\frac{1}{|\mathcal{D}_{test}|}\sum_{(g,v)\in \mathcal{D}_{test}}\frac{1}{rank(v)}.
\end{equation}
$MRR$ is an average of the reciprocal rank of the ground-truth item $v$ among all items except those which group $g$ has also interacted with, and
a good recommender model should have a bigger $MRR$ value.

Similarly, we also apply the above evaluation procedure to the personalized recommendation for individual users.
%

\subsection{Comparison Methods}

To answer the four research questions, we design the following four experiments with different comparison methods.


\textbf{Experiment 1} To answer RQ1, we compare our CAGR with a wide range of state-of-the-art group recommender models.

\textbf{SIGR}~\cite{yin2019social}: Social Influence-based Group Recommender is our previously proposed group recommendation model that leverages BGEM and the attention mechanism as building blocks  to learn both user embeddings and user social influences in a unified way.
SIGR uses the pure BGEM to model user embeddings from the user-item interaction data, and SIGR represents a group using a simple weighted sum aggregation function over raw member embeddings in a two-stage manner.

In~\cite{yin2019social},  SIGR has already been compared with the AGREE model~\cite{Cao:2018:AGR:3209978.3209998}, the PIT model~\cite{Liu:2012:EPI:2396761.2396848} and various simple aggregation strategies such as the simplest average strategy~\cite{Baltrunas:2010:GRR:1864708.1864733}, the least misery strategy~\cite{Amer-Yahia:2009:GRS:1687627.1687713} and  the maximum pleasure strategy~\cite{Baltrunas:2010:GRR:1864708.1864733}. SIGR has shown its superior performance over them.
Therefore, we do not compare CAGR with these models in this work, and we refer the interested readers to~\cite{yin2019social} for the results.

\textbf{GroupSA}~\cite{guo2020group}:
The Group Self-Attention model  treats the group decision making process as multiple voting processes, and develops a stacked social self-attention network to simulate how a group consensus is reached.
Based on the user-item and user-user interactions, GroupSA proposes two types of aggregation methods (i.e., item aggregation and social aggregation) to enhance the representation of users.

\textbf{SACML}~\cite{wang2019group}: 
The Self-Attention and Collaborative Metric Learning model employs the self-attention mechanism to automatically learn the weight of each group member, which are then aggregated to form a group. After that, the collaborative metric learning technique is leveraged to obtain the   group and item  representations in the embedding space.

\textbf{SA-NCF}~\cite{yang2020self}:
This method also employs the self-attention mechanism to learn the weight of each group member, and then learn the group and item embeddings via the Neural Collaborative Filter (NCF)~\cite{he2017neural} based on the group-item interaction data. 


%

\textbf{Experiment 2} To answer RQ2, we design three versions of our CAGR with different user embedding learning approaches: \textbf{CAGR-N}, \textbf{CAGR-G} and \textbf{CAGR-C}. CAGR-N does not use the GCNs-based method to learn user embeddings, i.e., CAGR only adopts the pure BGEM on the user-item interaction data. CAGR-G  uses the vanilla GCNs method GraphSAGE and CAGR-C uses the centrality-aware GraphCSC to learn user embeddings from both the user-item interaction and the user social network data.


\textbf{Experiment 3} To answer RQ3, we compare three model optimization approaches: Simple Training (\textbf{ST}), Two-stage Training (\textbf{TST})  and Joint Training (\textbf{JT}). \textbf{ST} optimizes our CAGR model only on the group-item interaction data, while \textbf{TST} and \textbf{JT} integrate the user activity data. For the joint training approach, we further compare two different negative sampling strategies: the \textbf{classic} degree-based sampling strategy and our proposed \textbf{group}-aware sampling strategy. Thus, we implement two versions of \textbf{JT}: \textbf{JT-C} and \textbf{JT-G}.

\textbf{Experiment 4} To answer RQ4, we investigate how the performance of our CAGR varies w.r.t. different key hyper-parameter setups, including  the model dimension $d$ that we set for all model embeddings such as the user and group embeddings, the number of parallel heads  $h$  in the multi-headed self-attention mechanism, the number of iterations $N$,  the number of negative samples $M$ and the number of neighbors to sample for any user $i$  $||\mathcal{N}_i||$ in Algorithm 1. For other hyper-parameters in the GraphCSC component, we follow their optimal setup in~\cite{chen2019exploiting,kipf2016semi}. For  other hyper-parameters in the model, we perform cross-validation and use the grid search algorithm to obtain the optimal hyper-parameter setup on the validation dataset.




\vspace{-5pt}
\section{Experimental Results}

In this section, we report and analyze the results of our four experiments.
%
%
%
%
\begin{figure}[t]
	\centering
	\begin{subfigure}{0.98\linewidth}
		\centering
		\includegraphics[scale = 0.35]{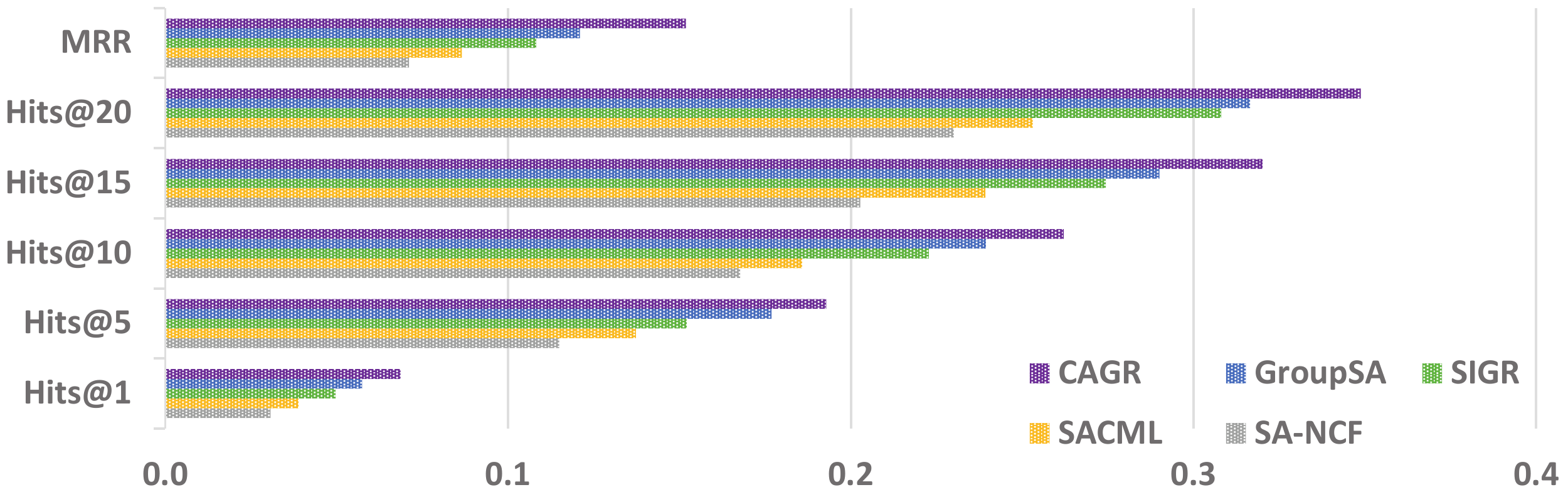}
		\caption{{\textbf{On Yelp Dataset}}}
	\end{subfigure}
	
	\begin{subfigure}{0.98\linewidth}
		\centering
		\includegraphics[scale = 0.35]{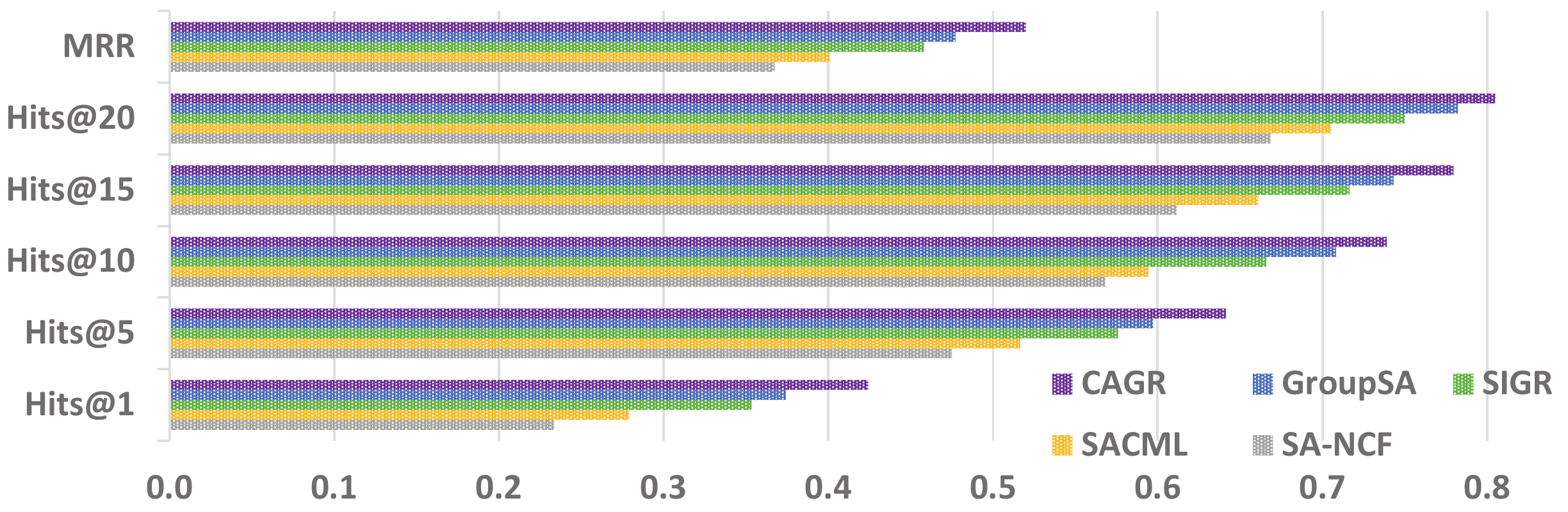}
		\caption{{\textbf{On Douban-Event Dataset}}}
	\end{subfigure}

	\begin{subfigure}{0.98\linewidth}
	\centering
	\includegraphics[scale = 0.35]{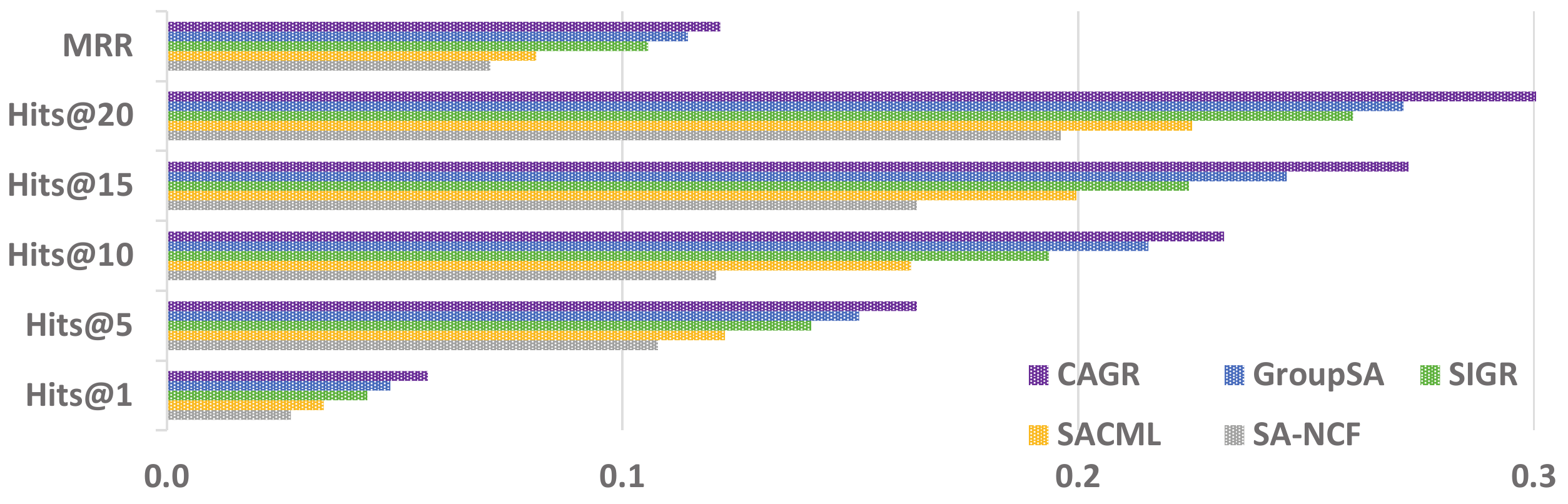}
	\caption{{\textbf{On Meetup Dataset}}}
\end{subfigure}

	\caption{{\textbf{Group Recommendation Performance}.}}
	\label{fg:exp4}

\end{figure}

\vspace{-8pt}
\subsection{Overall Group Recommendation Performance (RQ1)}

Fig.~\ref{fg:exp4} shows the results of Experiment 1 on Yelp, Douban-Event and Meetup datasets.
We have the following observations.
(1) Our proposed CAGR achieves the best performance on the three datasets for group recommendation, significantly outperforming other state-of-the-art group recommender models (all the $p$-values between our CAGR and each comparison method are much smaller than $0.01$, which indicates that the improvements are statistically significant). This validates the effectiveness of our CAGR solution.
In particular, CAGR outperforms SIGR, proving the superiority of our centrality-aware user representation learning  in capturing and integrating the centrality information from the social network to overcome the user data sparsity, and the end-to-end self-attention group representation method  in modeling the complex intra-group interactions.
CAGR also outperforms GroupSA, which validates the importance of taking the centrality information in social network for group representation learning.
(2) GroupSA achieves the second best performance, which validates that it is crucial to leverage the users' social information in group recommendation, and also demonstrates the effectiveness of the self-attention mechanism to learn user representations.
(3) SIGR outperforms SACML and SA-NCF in this experiment, and SIGR also outperforms many other baseline methods such as AGREE and PIT shown in~\cite{yin2019social}, which proves that it is important to exploit and integrate the user-item interaction data and the social network data to overcome the sparsity issue of the group-item interaction data in learning user embedding and personal social influence respectively.
This is in contrast with AGREE that cannot leverage the social network structure information and PIT that is incapable of integrating user-item interaction data.
(4) CAGR, GroupSA, SIGR, SACML and SA-NCF consistently outperform the simple group aggregation strategies BGEM+avg, BGEM+mp and BGEM+ml (see the results in ~\cite{yin2019social}), showing the advantage of automatically learning the aggregation strategy from data over the predefined aggregation strategies.

\begin{figure}[b]
	\begin{subfigure}{0.48\linewidth}
		\includegraphics[scale = 0.17]{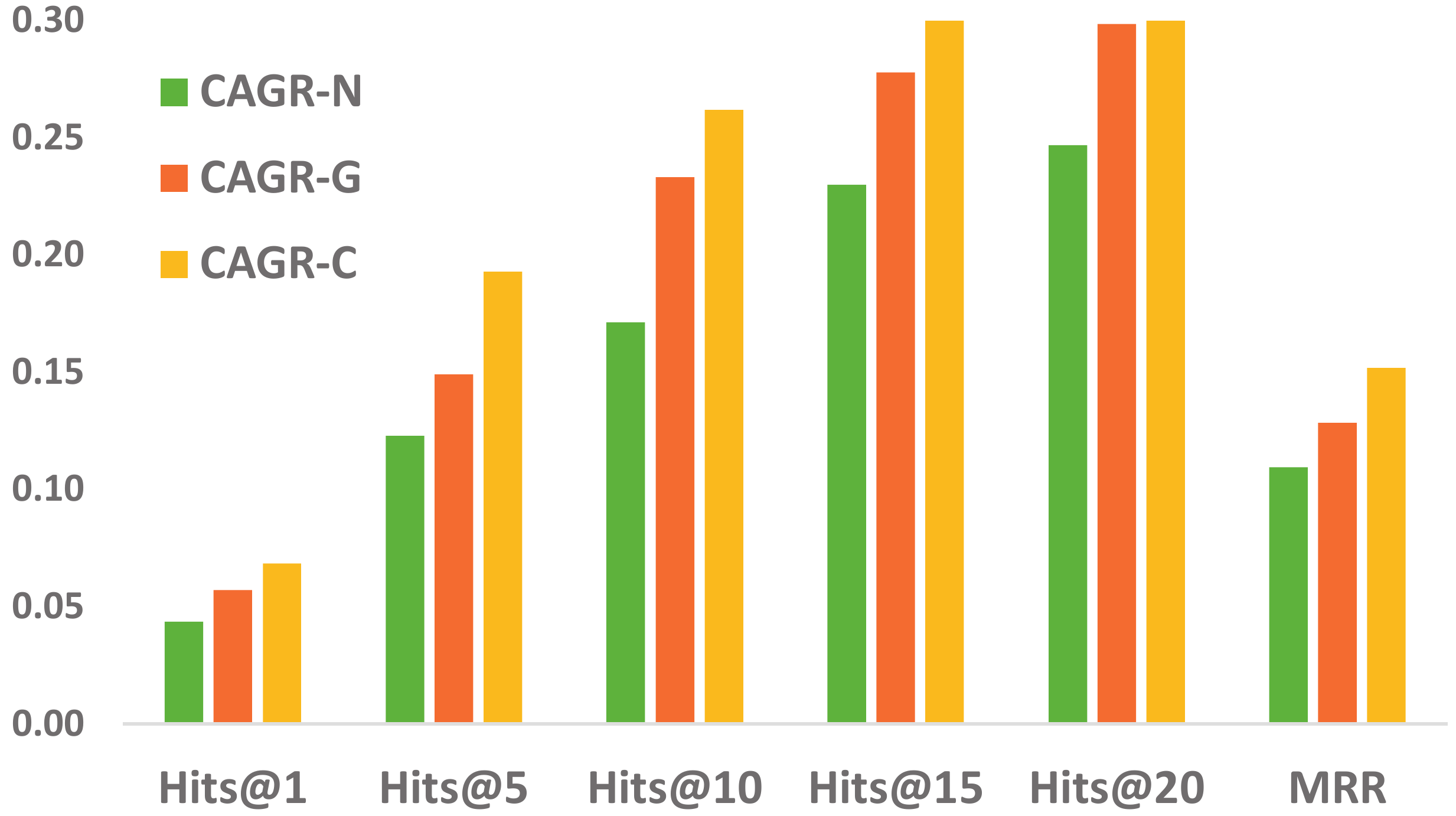}
		\caption{\textbf{On Yelp Dataset}}
	\end{subfigure}
	\hspace{0.1cm}
	\begin{subfigure}{0.49\linewidth}
		\includegraphics[scale = 0.17]{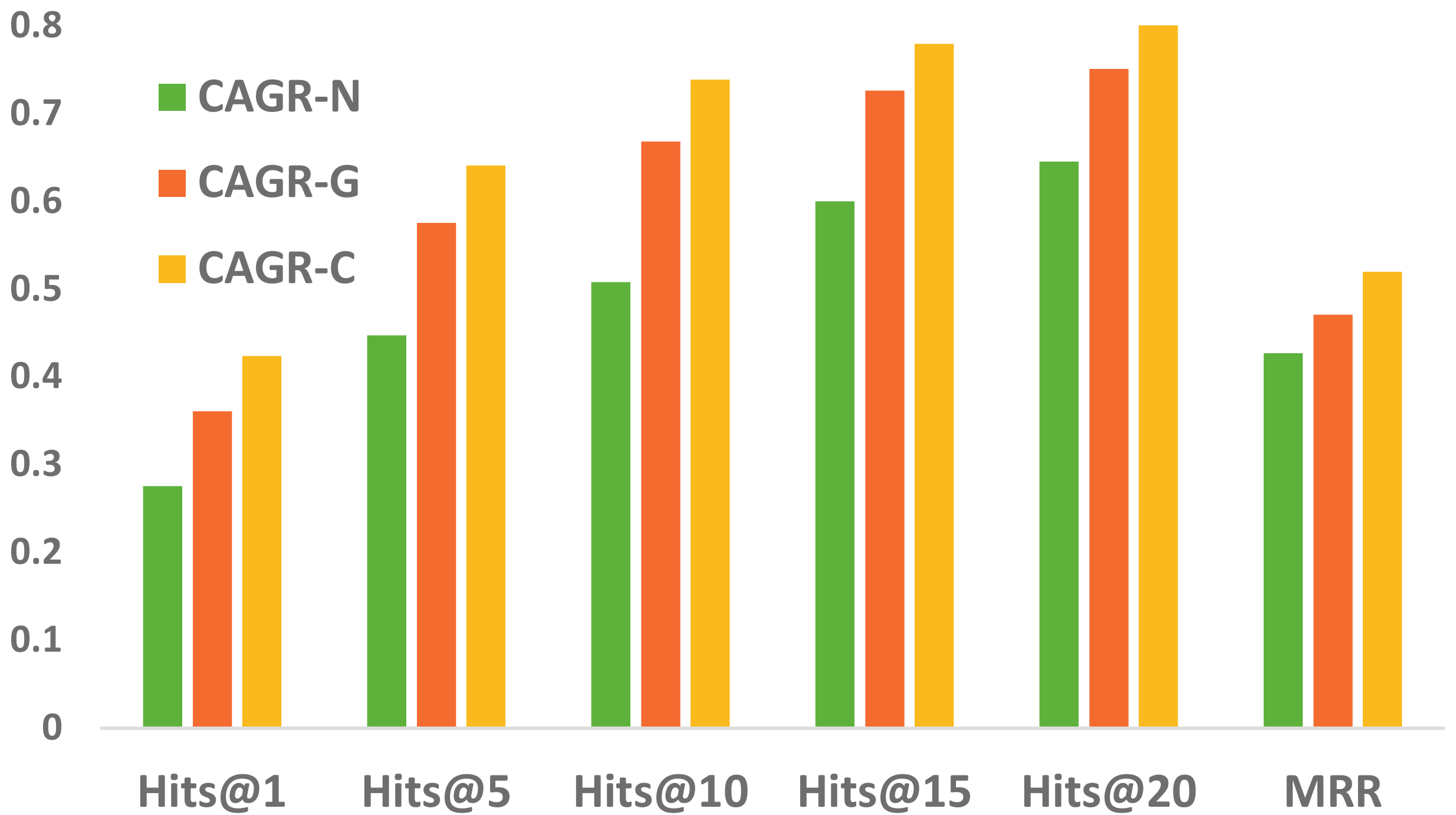}
		\caption{\textbf{On Douban-Event Dataset}}
	\end{subfigure}
	\caption{\textbf{Effect of Different Network Embedding Methods}.}
	\label{fg:exp2}
\end{figure}

\vspace{-5pt}
\subsection{Importance of User Embedding Techniques (RQ2)}
Fig.~\ref{fg:exp2} shows the results of Experiment 2 on Yelp and Douban-Event datasets.
The following observations are made from the results.
(1) Both CAGR-G and CAGR-C significantly and consistently outperform CAGR-N on both Yelp and Douban-Event datasets, showing that it is important to leverage the social network information to enhance the user embedding learning for group recommendation.
As the user-item interaction data is sparse, directly applying BGEM to estimate user embeddings from it would generate sub-optimal group recommendations.
On the other hand, the improvement by CAGR-G and CAGR-C demonstrates that integrating the social network data using GCNs can effectively alleviate the  user-item interaction data sparsity.
(2) CAGR-C achieves higher recommendation accuracy than CAGR-G, which validates that incorporating the centrality information into the user embeddings is helpful to model user social impacts and leads to more accurate group representations.



\vspace{-8pt}
\subsection{Importance of  User Activity Data (RQ3)}
\begin{figure}[!b]
	\begin{subfigure}{0.48\linewidth}
		\includegraphics[scale = 0.17]{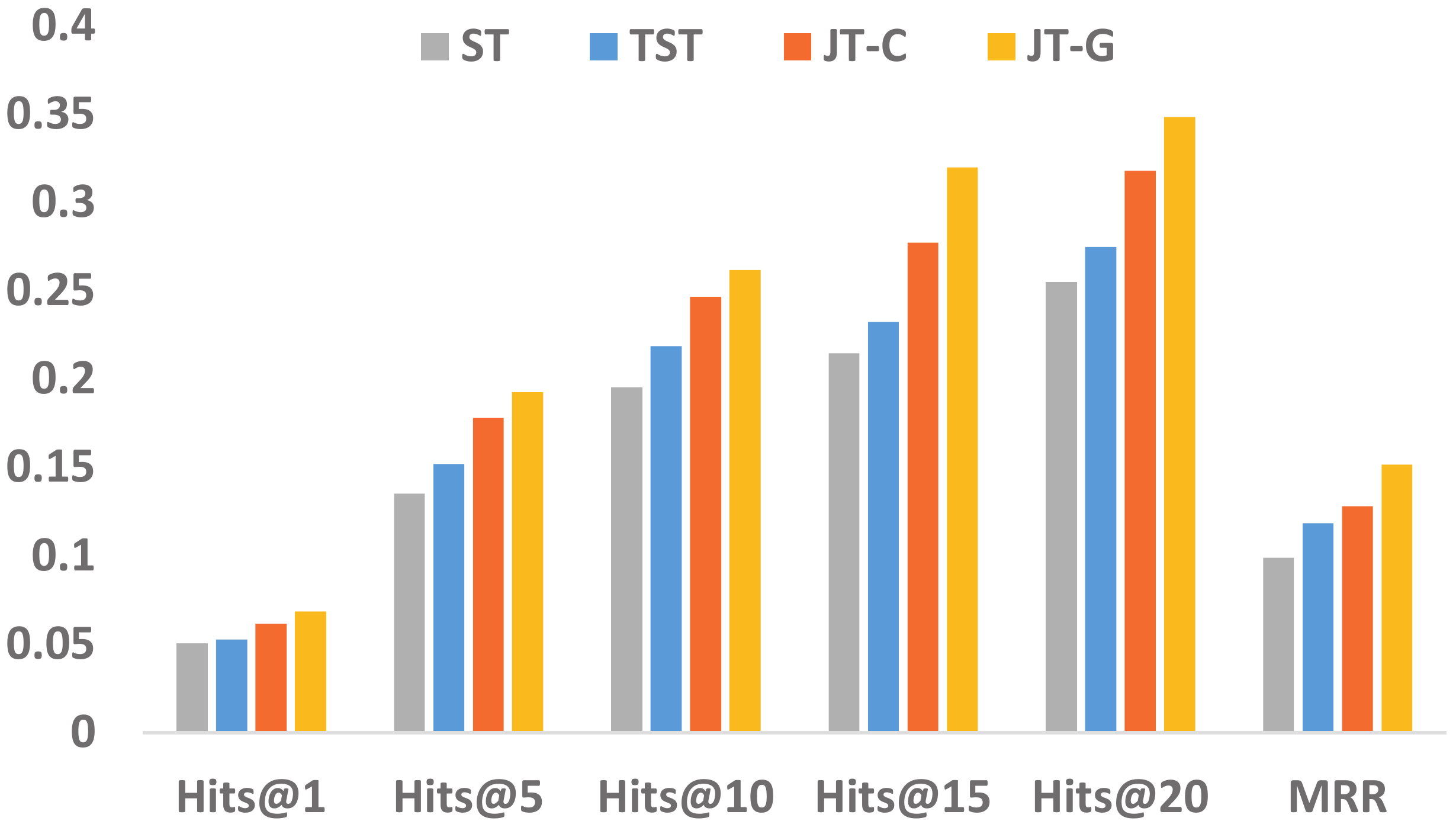}
		\caption{\textbf{On Yelp Dataset}}
	\end{subfigure}
	\hspace{0.1cm}
	\begin{subfigure}{0.49\linewidth}
		\includegraphics[scale = 0.17]{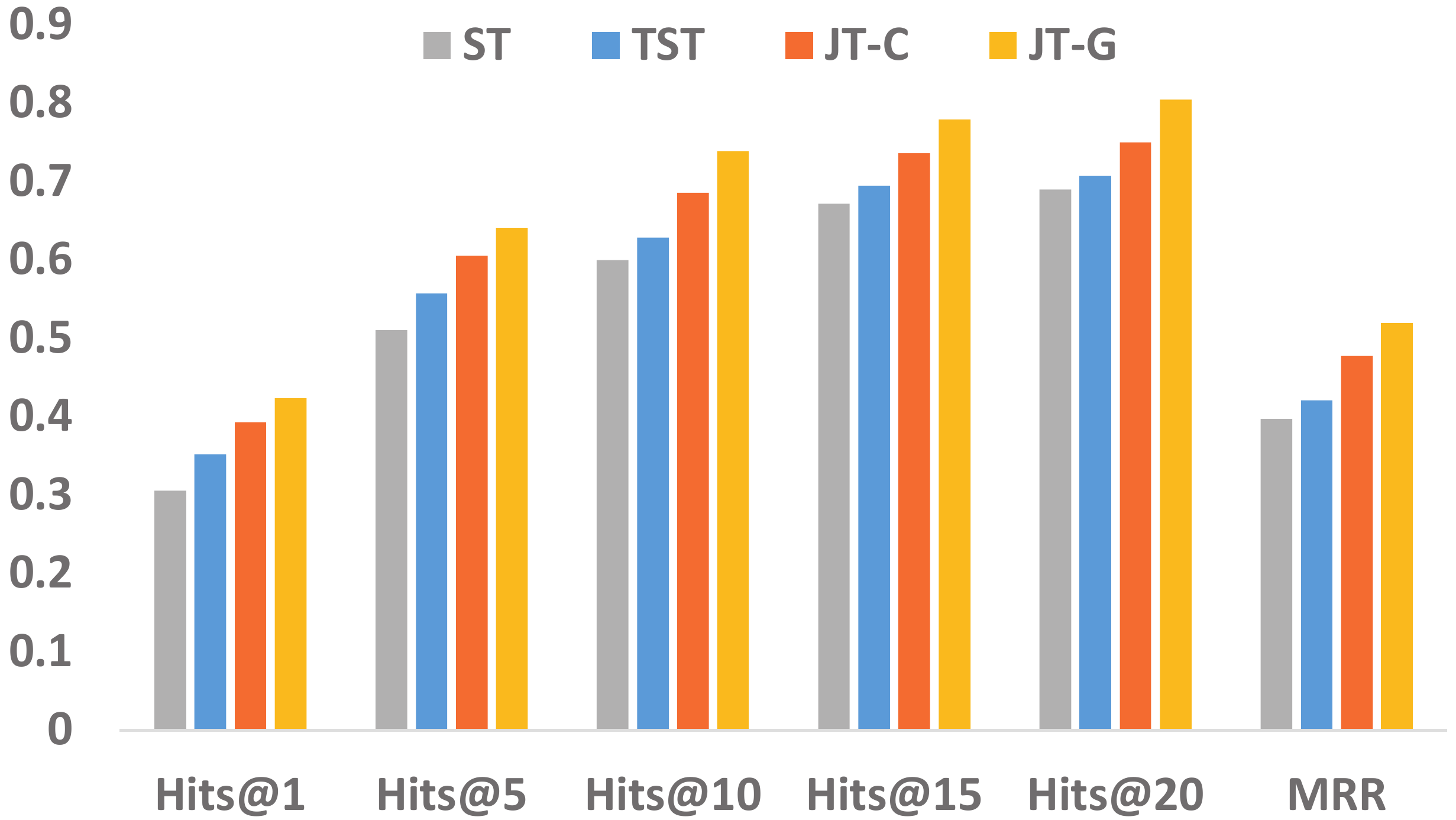}
		\caption{\textbf{On Douban-Event Dataset}}
	\end{subfigure}
	\caption{{\textbf{Comparison of Different Model Optimization Approaches}.}}
	\label{fg:exp3}
\end{figure}

Fig.~\ref{fg:exp3} shows the results of Experiment 3. We make three observations from the results. (1) All TST, JT-C and JT-G significantly outperform ST on both Yelp and Douban-Event datasets, showing the importance and necessity of leveraging the user activity data for training our CAGR model. (2) JT-C and JT-G perform better than TST, which shows the advantage of our proposed joint model optimization approach over the two-stage optimization approach.  This is because the embedding spaces separately learned from the user activity data and group-item interaction data may not be compatible. Besides, the experimental results also indicate that our proposed joint model optimization approach can effectively perform model optimization on heterogenous interaction data (e.g., the mixture of the user-item interaction data, user social network data and group-item interaction data) and address the issue of data skewness. (3) JT-G achieves better performance than JT-C, showing that our proposed group-aware negative sampling strategy is more suitable for the sparse group-item interaction graph than the classic degree-based negative sampling strategy.


\vspace{-4pt}
\subsection{Impact of Tuning Hyper-parameters (RQ4)}
Tables~\ref{tb:d} -- \ref{tb:nbrs} show the results of Experiment 4. Due to the space limitation, we only show the experimental results on Yelp dataset, and similar results are also achieved on Douban-Event dataset.

Both the size of model dimension $d$ (e.g., the size of user and group embeddings) and the number of heads $h$  are important to balance between the model size and the expressiveness. 
Moreover, the model can enjoy a higher degree of parallelization  with increasing $h$.
To study their impacts, we test the performance of CAGR  by varying the values of $d$ from 16 to 512 and varying the values of $h$ from 1 to 32.
Their  results are presented in Tables~\ref{tb:d} and Table~\ref{tb:h}.
From the results, we observe that CAGR achieves the best performance when $d=128$ and $h=16$. 
Another pattern we observe from both results is that the recommendation accuracy of CAGR first increases with the increasing $d$ (or $h$), and then begins to decrease.
The reason is that there may be excessive parameters in the model  if $d$ or $h$ is too large, which can introduce noise to the model and  lead to overfitting eventually.

Similarly, we investigate the converging performance of our CAGR model with increasing  number of iterations $N$ (i.e., the number of stochastic gradient steps) and  number of negative samples ($M$) drawn for each positive sample. Table~\ref{tb:iter} presents the performance of our CAGR model w.r.t. the number of iterations. When $N$ is larger than 4 millions, our model converges quickly and its performance becomes very stable. Table~\ref{tb:ns} shows the performance of CAGR w.r.t. the number of negative examples $M$. From the table, we observe that when the number of negative examples is larger than 6, the performance becomes very stable. Therefore, for each positive example, we do not need to sample many negative examples and just need to sample a few, which ensures the training efficiency of our CAGR model.

Finally, we study how the number of neighbors for each user $i$ in Algorithm 1  ($||\mathcal{N}_i||$) affects the model performance.
Specifically, we evaluate the model performance w.r.t. different settings of  $||\mathcal{N}_i||$ (i.e., 1, 2, 4, 6, 8 and 10) and then show the results in Table~\ref{tb:nbrs}.
From the experimental results, we observe that  our model achieves the peak results when $||\mathcal{N}_i|| = 4$, and then the results remain stable as the number increases.
This observation validates our intuition of the GraphCSC component that only a small number of important neighbors are sufficient to facilitate the learning of target node’s representation.

\begin{table}
	\footnotesize
	\centering
	\caption{{Impact of Parameter $d$ (the model dimension).}}
	\begin{tabular}{c|c|c|c|c|c|c}
$d$ & Hits@1 & Hits@5 & Hits@10 & Hits@15 & Hits@20 & MRR    \\ \hline
16  & 0.0496 & 0.1365 & 0.2101  & 0.2585  & 0.3093  & 0.1128 \\
32  & 0.0576 & 0.1738 & 0.2343  & 0.2731  & 0.3187  & 0.1244 \\
64  & 0.0625 & 0.1954 & 0.2497  & 0.3054  & 0.3348  & 0.1425 \\
128 & 0.0685 & 0.1989 & 0.2681  & 0.3261  & 0.3439  & 0.1578 \\
256 & 0.0636 & 0.1849 & 0.2411  & 0.3201  & 0.3398  & 0.1551 \\
512 & 0.0588 & 0.1827 & 0.2337  & 0.3135  & 0.3305  & 0.1377 \\ \hline
	\end{tabular}
	\label{tb:d}
\end{table}

\begin{table}[!t]
	\footnotesize
	\caption{{Impact of Parameter $h$ (the number of heads).}}
	\centering
	\begin{tabular}{c|c|c|c|c|c|c}
		
$h$ & Hits@1 & Hits@5 & Hits@10 & Hits@15 & Hits@20 & MRR    \\  \hline
1   & 0.0562 & 0.188  & 0.2488  & 0.3119  & 0.3319  & 0.1385 \\
2   & 0.0578 & 0.1909 & 0.2597  & 0.3159  & 0.3365  & 0.1439 \\
4   & 0.062  & 0.1945 & 0.2615  & 0.3175  & 0.3443  & 0.1505 \\
8   & 0.0659 & 0.1983 & 0.264   & 0.3222  & 0.3475  & 0.1495 \\
16  & 0.0685 & 0.1989 & 0.2681  & 0.3261  & 0.3547  & 0.1578 \\
32  & 0.0677 & 0.1956 & 0.2671  & 0.3255  & 0.3509  & 0.1556          \\ \hline
		
	\end{tabular}
	\label{tb:h}
\end{table}

\begin{table}[!t]
	\footnotesize
	\caption{{Impact of Parameter $N$ (the number of iterations).}}
	\centering
	\begin{tabular}{c|c|c|c|c|c|c}
	
$N$  & Hits@1 & Hits@5 & Hits@10 & Hits@15 & Hits@20 & MRR    \\ \hline
1m & 0.0652 & 0.1886 & 0.2592  & 0.317   & 0.3318  & 0.1493 \\
2m & 0.0664 & 0.1935 & 0.2606  & 0.3196  & 0.3463  & 0.1503 \\
3m & 0.0679 & 0.1972 & 0.2637  & 0.3211  & 0.3489  & 0.1521 \\
4m & 0.0685 & 0.1989 & 0.2681  & 0.3261  & 0.3547  & 0.1541 \\
5m & 0.0685 & 0.1989 & 0.2681  & 0.3261  & 0.3547  & 0.1578 \\
6m & 0.0683 & 0.1989 & 0.2681  & 0.3261  & 0.3547  & 0.1578 \\ \hline
	
\end{tabular}
	\label{tb:iter}
\end{table}

\begin{table}[!t]
	\footnotesize
	\caption{{Impact of Parameter $M$ (the number of negative edges sampled for each positive edge).}}
	\centering
	\begin{tabular}{c|c|c|c|c|c|c}

$M$ & Hits@1 & Hits@5 & Hits@10 & Hits@15 & Hits@20 & MRR    \\ \hline
2 & 0.0578 & 0.1876 & 0.261   & 0.3083  & 0.3333  & 0.1451 \\
3 & 0.0609 & 0.1908 & 0.2632  & 0.3151  & 0.3399  & 0.1518 \\
4 & 0.063  & 0.1927 & 0.2668  & 0.3183  & 0.346   & 0.1529 \\
5 & 0.0659 & 0.1954 & 0.2677  & 0.3233  & 0.3512  & 0.1549 \\
6 & 0.0685 & 0.1989 & 0.2681  & 0.3261  & 0.3547  & 0.1578 \\
7 & 0.0688 & 0.1991 & 0.2681  & 0.3262  & 0.3549  & 0.1579 \\ \hline
\end{tabular}
	\label{tb:ns}
\end{table}

\begin{table}[!t]
	\footnotesize
	\caption{{Impact of Parameter $||\mathcal{N}_i||$ (the number of neighbors to sample for any user $i$ in Algorithm 1).}}
	\centering
	\begin{tabular}{c|c|c|c|c|c|c}

		$||\mathcal{N}_i||$ & Hits@1 & Hits@5 & Hits@10 & Hits@15 & Hits@20 & MRR    \\ \hline
1                   & 0.053  & 0.175  & 0.249   & 0.309   & 0.336   & 0.143 \\
2                   & 0.062  & 0.188  & 0.257   & 0.317   & 0.341   & 0.147 \\
4                   & 0.068  & 0.192  & 0.262   & 0.320   & 0.348   & 0.151 \\
6                   & 0.068  & 0.192  & 0.262   & 0.320   & 0.348   & 0.151 \\
8                   & 0.069  & 0.191  & 0.262   & 0.321   & 0.348   & 0.151 \\
10                  & 0.068  & 0.191  & 0.261   & 0.319   & 0.347   & 0.151 \\ \hline
	\end{tabular}
	\label{tb:nbrs}
\end{table}

\vspace{-8pt}
\section{Related Work}



There are two lines of research on group recommendation based on the group types~\cite{Quintarelli:2016:RNI:2959100.2959137}. Groups with stable members and rich historical interactions are often referred as persistent groups (also called established groups) while groups formed by users ad-hoc are dubbed as occasional groups.  As a persistent group can be treated as a virtual user, conventional  personalized recommendation techniques  can be straightforwardly adopted for making recommendation to persistent groups~\cite{Hu:2014:DMG:2892753.2892811}. In this paper, we focus on making recommendations to occasional groups.

Making recommendations to occasional groups is much more challenging due to the lack of sufficient group-item interactions.  Existing studies on occasional group recommendations focus on aggregation approaches that aggregate individual preferences or recommendation results of the group members as the group preferences or group recommendations.  All these aggregation-based group recommendation approaches can be divided into two categories: \textbf{late aggregation} and \textbf{early aggregation}.

The late aggregation-based approaches~\cite{Xiao:2017:FGR:3109859.3109887,Amer-Yahia:2009:GRS:1687627.1687713,Salehi-Abari:2015:PSN:2792838.2800190} first generate the recommendation results or lists for each group member, and then aggregate these individual recommendation results to produce the group recommendations.  A variety of aggregation strategies~\cite{Amer-Yahia:2009:GRS:1687627.1687713,Baltrunas:2010:GRR:1864708.1864733,Quintarelli:2016:RNI:2959100.2959137,roy2014exploiting,Salehi-Abari:2015:PSN:2792838.2800190} have been proposed, such as average satisfaction, least misery and maximum pleasure, and most of them come from the social choice theory~\cite{brandt2012computational}. For example, average satisfaction assigns equal importance to each group member and assumes that each group member contributes equally to the group decision-making. However, these aggregation strategies are heuristic and manually predefined rather than data-driven or learned from data. \cite{Pessemier:2014:CGR:2662458.2662495} does a systematic evaluation of all existing predefined aggregation strategies. Their conclusion is that the best-performing aggregation strategy does not exist and their performances depend on the datasets. In other words, there does not exist a predefined fixed aggregation strategy which can perform best on all datasets.

In contrast, the early aggregation-based approaches, such as~\cite{Yuan:2014:CGM:2623330.2623616,Ye:2012:ESI:2348283.2348373,Cao:2018:AGR:3209978.3209998}, first aggregate either explicit or implicit user profiles  into a group profile or representation, and then produce the group recommendations based on the group profile or representation. The explicit user profiles refer to users' interaction records on items, and the implicit user profiles refer to the latent representations of users' preferences, such as user embedding.
A line of this type of work is based on probabilistic generative models or more precisely topic models~\cite{Liu:2012:EPI:2396761.2396848,Yuan:2014:CGM:2623330.2623616,7498303,7491346}, which model groups by capturing both personal preferences of group members and their impacts in the group. The basic assumption of these models is that users should be treated differently and the notion of \textit{influence} is introduced to quantify the contribution of each group member to the group decision making and implement the data-driven aggregation.
Although these models share the similar intuitions with our proposed CAGR model, they do not consider the sparsity issue of the group-item interaction data.
Moreover, our CAGR model is technically different from them such as PIT and COM. Compared with our proposed CAGR that takes the bipartite graph embedding model BGEM as the foundation, these topic model-based group recommender models have limited modeling and expressive abilities, since they constrain the key parameters (e.g., users' personal preferences) to be a probability distribution. Moreover, these models are not as flexible as our CAGR,  and they are incapable of seamlessly integrating the user activity data to improve the estimation of users' personal preferences.

Recently, \cite{Cao:2018:AGR:3209978.3209998} developed an Attentive Group Recommendation system (AGREE) by combining a standard attention network with the neural collaborative filtering method (NCF). Compared with our CAGR, AGREE has two serious drawbacks.  First, it does not consider the data sparsity issue of the group-item interaction data in learning user weights. Second, its good performance heavily depends on the direct learning of group preference embedding from the group's interaction data and cannot make good recommendations for cold-start groups without any interaction record.

In our previous work~\cite{yin2019social}, we proposed Social Influence-based Group Recommender (SIGR) that represents groups by aggregating the members' preferences based on the social influences.
Compared with SIGR, CAGR makes the following significant improvements. 
\textbf{End-to-end:} SIGR is a two-stage method that cannot be trained in an end-to-end manner,
and CAGR addresses this limitation by introducing a single optimization criterion to enable end-to-end training.
\textbf{Expressiveness:} SIGR uses the plain linear weighted sum to aggregate raw user embeddings to represent the group, but it ignores the complex interactions among them. To address this problem, our CAGR handles this problem with a self-attention mechanism that can better model the intra-relations in the group.
\textbf{Mitigating user data sparsity:} SIGR learns user embeddings  using the pure BGEM framework from the user-item interaction data, which is too sparse to support accurate estimation of user representations.
CAGR mitigates the user data sparsity problem and enhances the user embedding learning  by additionally considering  the user social network data and incorporating the node centrality information in the network.

Our work can be categorized as  early aggregation-based approaches and we focus on overcoming the sparsity issue and limitations of the group-item  and user-item interaction data in learning both users' personal and groups' preferences.

\vspace{-5pt}
\section{Conclusions and Future Work}
In this paper, we focused on the problem of making recommendations to occasional groups.
We proposed an end-to-end Centrality-Aware Group Recommender (CAGR) model to enhance the group and user representation learning by overcoming the bottleneck data sparsity problems.
Specifically,  to mitigate the sparsity of data generated by occasional groups,
we proposed to represent a group by aggregating its member representations based on a novel and end-to-end self-attention mechanism.
To learn centrality-aware user embeddings, we developed a graph convolution operation on the user social network data in order to overcome the user-item interaction data sparsity problem.
We also proposed two model optimization approaches to leverage the user embedding learning procedure to help improve the parameters estimation.
To evaluate the performance of group recommender systems in making recommendations to occasional groups, we created three large-scale benchmark datasets and conducted extensive experiments on them. The experimental results validated the superiority of our solutions by comparing with the state-of-the-art techniques.

Having alleviated the data sparsity problem in group recommendation to a great extent, we plan to further investigate another rarely explored  problem in group recommendation: the propensity problem~\cite{wang2019doubly}.
This problem arises because the probabilities  of different ratings  being observed in most datasets are not the same.
More specifically,  users tend to only rate an item that they like and thus the ratings of a lower value are more likely to be missing~\cite{lim2015top}.
This type of bias commonly existing in the dataset would lead to  inaccurate user preference modeling results, and hence eventually harm the group representation learning.
To this end, in the future, we intend to employ the state-of-the-art techniques, such as~\cite{wang2019doubly}, in our CAGR model to address the propensity problem and further improve the group recommendation performance.

\vspace{-5pt}
\section{Acknowledgement}
This work was supported by ARC Discovery Project (Grant No.DP190101985, DP170103954)  and  National Natural Science Foundation of China  (Grant No.61532018, 61836007, 61832017, 61632016).

\bibliographystyle{IEEEtran}  
\bibliography{bibi}  

\begin{thebibliography}{10}
\providecommand{\url}[1]{#1}
\csname url@samestyle\endcsname
\providecommand{\newblock}{\relax}
\providecommand{\bibinfo}[2]{#2}
\providecommand{\BIBentrySTDinterwordspacing}{\spaceskip=0pt\relax}
\providecommand{\BIBentryALTinterwordstretchfactor}{4}
\providecommand{\BIBentryALTinterwordspacing}{\spaceskip=\fontdimen2\font plus
\BIBentryALTinterwordstretchfactor\fontdimen3\font minus
  \fontdimen4\font\relax}
\providecommand{\BIBforeignlanguage}[2]{{%
\expandafter\ifx\csname l@#1\endcsname\relax
\typeout{** WARNING: IEEEtran.bst: No hyphenation pattern has been}%
\typeout{** loaded for the language `#1'. Using the pattern for}%
\typeout{** the default language instead.}%
\else
\language=\csname l@#1\endcsname
\fi
#2}}
\providecommand{\BIBdecl}{\relax}
\BIBdecl

\bibitem{McCarthy:1998:MAG:289444.289511}
J.~F. McCarthy and T.~D. Anagnost, ``Musicfx: An arbiter of group preferences
  for computer supported collaborative workouts,'' in \emph{CSCW}, 1998, pp.
  363--372.

\bibitem{mccarthy2006cats}
K.~McCarthy, M.~Salam{\'o}, L.~Coyle, L.~McGinty, B.~Smyth, and P.~Nixon,
  ``Cats: A synchronous approach to collaborative group recommendation,'' in
  \emph{FLAIRS}, 2006, pp. 86--91.

\bibitem{Liu:2012:EPI:2396761.2396848}
X.~Liu, Y.~Tian, M.~Ye, and W.-C. Lee, ``Exploring personal impact for group
  recommendation,'' in \emph{CIKM}, 2012, pp. 674--683.

\bibitem{Quintarelli:2016:RNI:2959100.2959137}
E.~Quintarelli, E.~Rabosio, and L.~Tanca, ``Recommending new items to ephemeral
  groups using contextual user influence,'' in \emph{RecSys}, 2016, pp.
  285--292.

\bibitem{Xiao:2017:FGR:3109859.3109887}
L.~Xiao, Z.~Min, Z.~Yongfeng, G.~Zhaoquan, L.~Yiqun, and M.~Shaoping,
  ``Fairness-aware group recommendation with pareto-efficiency,'' in
  \emph{RecSys}, 2017, pp. 107--115.

\bibitem{Hu:2014:DMG:2892753.2892811}
L.~Hu, J.~Cao, G.~Xu, L.~Cao, Z.~Gu, and W.~Cao, ``Deep modeling of group
  preferences for group-based recommendation,'' in \emph{AAAI}, 2014, pp.
  1861--1867.

\bibitem{Said:2011:GRC:2096112.2096113}
A.~Said, S.~Berkovsky, and E.~W. De~Luca, ``Group recommendation in context,''
  in \emph{CAMRa}, 2011, pp. 2--4.

\bibitem{Ronen:2014:RSM:2600428.2609596}
I.~Ronen, I.~Guy, E.~Kravi, and M.~Barnea, ``Recommending social media content
  to community owners,'' in \emph{SIGIR}, 2014, pp. 243--252.

\bibitem{Cao:2018:AGR:3209978.3209998}
D.~Cao, X.~He, L.~Miao, Y.~An, C.~Yang, and R.~Hong, ``Attentive group
  recommendation,'' in \emph{SIGIR}, 2018, pp. 645--654.

\bibitem{Baltrunas:2010:GRR:1864708.1864733}
L.~Baltrunas, T.~Makcinskas, and F.~Ricci, ``Group recommendations with rank
  aggregation and collaborative filtering,'' in \emph{RecSys}, 2010, pp.
  119--126.

\bibitem{Yuan:2014:CGM:2623330.2623616}
Q.~Yuan, G.~Cong, and C.-Y. Lin, ``Com: A generative model for group
  recommendation,'' in \emph{KDD}, 2014, pp. 163--172.

\bibitem{Amer-Yahia:2009:GRS:1687627.1687713}
S.~Amer-Yahia, S.~B. Roy, A.~Chawlat, G.~Das, and C.~Yu, ``Group
  recommendation: Semantics and efficiency,'' \emph{Proc. VLDB Endow.}, vol.~2,
  no.~1, pp. 754--765, Aug. 2009.

\bibitem{Salehi-Abari:2015:PSN:2792838.2800190}
A.~Salehi-Abari and C.~Boutilier, ``Preference-oriented social networks: Group
  recommendation and inference,'' in \emph{RecSys}, 2015, pp. 35--42.

\bibitem{yin2019social}
H.~Yin, Q.~Wang, K.~Zheng, Z.~Li, J.~Yang, and X.~Zhou, ``Social
  influence-based group representation learning for group recommendation,'' in
  \emph{ICDE}, 2019, pp. 566--577.

\bibitem{vaswani2017attention}
A.~Vaswani, N.~Shazeer, N.~Parmar, J.~Uszkoreit, L.~Jones, A.~N. Gomez,
  {\L}.~Kaiser, and I.~Polosukhin, ``Attention is all you need,'' in
  \emph{Advances in neural information processing systems}, 2017, pp.
  5998--6008.

\bibitem{zafarani2014social}
R.~Zafarani, M.~A. Abbasi, and H.~Liu, \emph{Social Media Mining: an
  Introduction}.\hskip 1em plus 0.5em minus 0.4em\relax Cambridge University
  Press, 2014.

\bibitem{mcpherson2001birds}
M.~McPherson, L.~Smith-Lovin, and J.~M. Cook, ``Birds of a feather: Homophily
  in social networks,'' \emph{Annual review of sociology}, pp. 415--444, 2001.

\bibitem{yin2018joint}
H.~Yin, L.~Zou, Q.~V.~H. Nguyen, Z.~Huang, and X.~Zhou, ``Joint event-partner
  recommendation in event-based social networks,'' in \emph{ICDE}, 2018, pp.
  929--940.

\bibitem{DBLP:conf/nips/MikolovSCCD13}
T.~Mikolov, I.~Sutskever, K.~Chen, G.~S. Corrado, and J.~Dean, ``Distributed
  representations of words and phrases and their compositionality,'' in
  \emph{NIPS}, 2013, pp. 3111--3119.

\bibitem{boratto2010state}
L.~Boratto and S.~Carta, ``State-of-the-art in group recommendation and new
  approaches for automatic identification of groups,'' in \emph{Information
  retrieval and mining in distributed environments}, 2010, pp. 1--20.

\bibitem{7436655}
J.~Guo, Y.~Zhu, A.~Li, Q.~Wang, and W.~Han, ``A social influence approach for
  group user modeling in group recommendation systems,'' \emph{IEEE Intelligent
  Systems}, vol.~31, no.~5, pp. 40--48, Sept 2016.

\bibitem{alina2014social}
I.~Alina~Christensen and S.~Schiaffino, ``Social influence in group recommender
  systems,'' \emph{Online Information Review}, vol.~38, no.~4, pp. 524--542,
  2014.

\bibitem{bahdanau2014neural}
D.~Bahdanau, K.~Cho, and Y.~Bengio, ``Neural machine translation by jointly
  learning to align and translate,'' \emph{arXiv preprint arXiv:1409.0473},
  2014.

\bibitem{Perozzi:2014:DOL:2623330.2623732}
B.~Perozzi, R.~Al-Rfou, and S.~Skiena, ``Deepwalk: Online learning of social
  representations,'' in \emph{KDD}, 2014, pp. 701--710.

\bibitem{Grover:2016:NSF:2939672.2939754}
A.~Grover and J.~Leskovec, ``Node2vec: Scalable feature learning for
  networks,'' in \emph{KDD}, 2016, pp. 855--864.

\bibitem{parikh2016decomposable}
A.~Parikh, O.~Tackstrom, D.~Das, and J.~Uszkoreit, ``A decomposable attention
  model for natural language inference,'' in \emph{EMNLP}, 2016, pp.
  2249--2255.

\bibitem{yang2016hierarchical}
Z.~Yang, D.~Yang, C.~Dyer, X.~He, A.~Smola, and E.~Hovy, ``Hierarchical
  attention networks for document classification,'' in \emph{NAACL-HLT}, 2016,
  pp. 1480--1489.

\bibitem{newman2002assortative}
M.~E. Newman, ``Assortative mixing in networks,'' \emph{Physical review
  letters}, vol.~89, no.~20, p. 208701, 2002.

\bibitem{chen2019exploiting}
H.~Chen, H.~Yin, T.~Chen, Q.~V.~H. Nguyen, W.-C. Peng, and X.~Li, ``Exploiting
  centrality information with graph convolutions for network representation
  learning,'' in \emph{ICDE}, 2019, pp. 590--601.

\bibitem{nair2010rectified}
V.~Nair and G.~E. Hinton, ``Rectified linear units improve restricted boltzmann
  machines,'' in \emph{ICML}, 2010, pp. 807--814.

\bibitem{krizhevsky2012imagenet}
A.~Krizhevsky, I.~Sutskever, and G.~E. Hinton, ``Imagenet classification with
  deep convolutional neural networks,'' in \emph{NIPS}, 2012, pp. 1097--1105.

\bibitem{kipf2016semi}
T.~N. Kipf and M.~Welling, ``Semi-supervised classification with graph
  convolutional networks,'' \emph{arXiv}, 2016.

\bibitem{page1999pagerank}
L.~Page, S.~Brin, R.~Motwani, and T.~Winograd, ``The pagerank citation ranking:
  Bringing order to the web,'' Stanford InfoLab, Tech. Rep., 1999.

\bibitem{Xiemincikm:2016}
M.~Xie, H.~Yin, H.~Wang, and F.~Xu, ``Learning graph-based poi embedding for
  location-based recommendation,'' in \emph{CIKM}, 2016.

\bibitem{rendle2009bpr}
S.~Rendle, C.~Freudenthaler, Z.~Gantner, and L.~Schmidt-Thieme, ``Bpr: Bayesian
  personalized ranking from implicit feedback,'' in \emph{UAI}, 2009, pp.
  452--461.

\bibitem{he2016fast}
X.~He, H.~Zhang, M.-Y. Kan, and T.-S. Chua, ``Fast matrix factorization for
  online recommendation with implicit feedback,'' in \emph{SIGIR}, 2016, pp.
  549--558.

\bibitem{he2017neural}
X.~He, L.~Liao, H.~Zhang, L.~Nie, X.~Hu, and T.-S. Chua, ``Neural collaborative
  filtering,'' in \emph{WWW}, 2017, pp. 173--182.

\bibitem{liu2012event}
X.~Liu, Q.~He, Y.~Tian, W.-C. Lee, J.~McPherson, and J.~Han, ``Event-based
  social networks: linking the online and offline social worlds,'' in
  \emph{SIGKDD}, 2012, pp. 1032--1040.

\bibitem{cremonesi2010performance}
P.~Cremonesi, Y.~Koren, and R.~Turrin, ``Performance of recommender algorithms
  on top-n recommendation tasks,'' in \emph{RecSys}, 2010, pp. 39--46.

\bibitem{hu2013spatial}
B.~Hu and M.~Ester, ``Spatial topic modeling in online social media for
  location recommendation,'' in \emph{RecSys}, 2013, pp. 25--32.

\bibitem{Wang2018}
Q.~Wang, H.~Yin, Z.~Hu, D.~Lian, H.~Wang, and Z.~Huang, ``Neural memory
  streaming recommender networks with adversarial training,'' in \emph{KDD},
  2018, pp. 2467--2475.

\bibitem{guo2020group}
L.~Guo, H.~Yin, Q.~Wang, B.~Cui, Z.~Huang, and L.~Cui, ``Group recommendation
  with latent voting mechanism,'' in \emph{ICDE}, 2020, pp. 121--132.

\bibitem{wang2019group}
H.~Wang, Y.~Li, and F.~Frimpong, ``Group recommendation via self-attention and
  collaborative metric learning model,'' \emph{IEEE Access}, vol.~7, pp.
  164\,844--164\,855, 2019.

\bibitem{yang2020self}
X.~Yang and Y.~Shi, ``Self-attention-based group recommendation,'' in
  \emph{ITNEC}, vol.~1, 2020, pp. 2540--2546.

\bibitem{roy2014exploiting}
S.~B. Roy, S.~Thirumuruganathan, S.~Amer-Yahia, G.~Das, and C.~Yu, ``Exploiting
  group recommendation functions for flexible preferences,'' in \emph{ICDE},
  2014, pp. 412--423.

\bibitem{brandt2012computational}
F.~Brandt, V.~Conitzer, and U.~Endriss, ``Computational social choice,''
  \emph{Multiagent systems}, pp. 213--283, 2012.

\bibitem{Pessemier:2014:CGR:2662458.2662495}
T.~Pessemier, S.~Dooms, and L.~Martens, ``Comparison of group recommendation
  algorithms,'' \emph{Multimedia Tools Appl.}, vol.~72, no.~3, pp. 2497--2541,
  Oct. 2014.

\bibitem{Ye:2012:ESI:2348283.2348373}
M.~Ye, X.~Liu, and W.-C. Lee, ``Exploring social influence for recommendation:
  A generative model approach,'' in \emph{SIGIR}, 2012, pp. 671--680.

\bibitem{7498303}
H.~Yin, Z.~Hu, X.~Zhou, H.~Wang, K.~Zheng, Q.~V.~H. Nguyen, and S.~Sadiq,
  ``Discovering interpretable geo-social communities for user behavior
  prediction,'' in \emph{ICDE}, 2016, pp. 942--953.

\bibitem{7491346}
H.~Yin, X.~Zhou, B.~Cui, H.~Wang, K.~Zheng, and Q.~V.~H. Nguyen, ``Adapting to
  user interest drift for poi recommendation,'' \emph{TKDE}, vol.~28, no.~10,
  pp. 2566--2581, 2016.

\bibitem{wang2019doubly}
X.~Wang, R.~Zhang, Y.~Sun, and J.~Qi, ``Doubly robust joint learning for
  recommendation on data missing not at random,'' in \emph{ICML}, 2019, pp.
  6638--6647.

\bibitem{lim2015top}
D.~Lim, J.~McAuley, and G.~Lanckriet, ``Top-n recommendation with missing
  implicit feedback,'' in \emph{Proceedings of the 9th ACM Conference on
  Recommender Systems}, 2015, pp. 309--312.

\end{thebibliography}
\vskip -2\baselineskip plus -1fil
\begin{IEEEbiography}[{\includegraphics[width=1in,height=1.15in,clip,keepaspectratio]{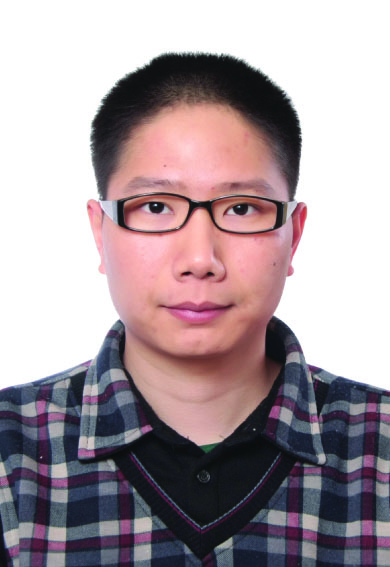}}]{Hongzhi Yin} works as a senior lecturer  with The University of Queensland, Australia. He received his doctoral degree from Peking University in July 2014, and his PhD Thesis won the highly competitive Distinguished Doctor Degree Thesis Award of Peking University. His current main research interests include recommender system, social media analytics and mining, network embedding and mining, time series data and sequence data mining and learning, chatbots, federated learning, topic models, deep learning and smart transportation.
He has published 120+ papers in top conferences and journals.
He has won ARC Discovery Early Career Researcher Award 2016 and  UQ Foundation Research Excellence Award 2019 (the first winner of this award in his school since the establishment of this award 20 years ago) as well as 5 Best Paper Awards such as ICDE'19 Best Paper Award and ACM Annual Best Computing Award.
\end{IEEEbiography}
\vskip -2\baselineskip plus -1fil
\begin{IEEEbiography}[{\includegraphics[width=1in,height=1.15in,clip,keepaspectratio]{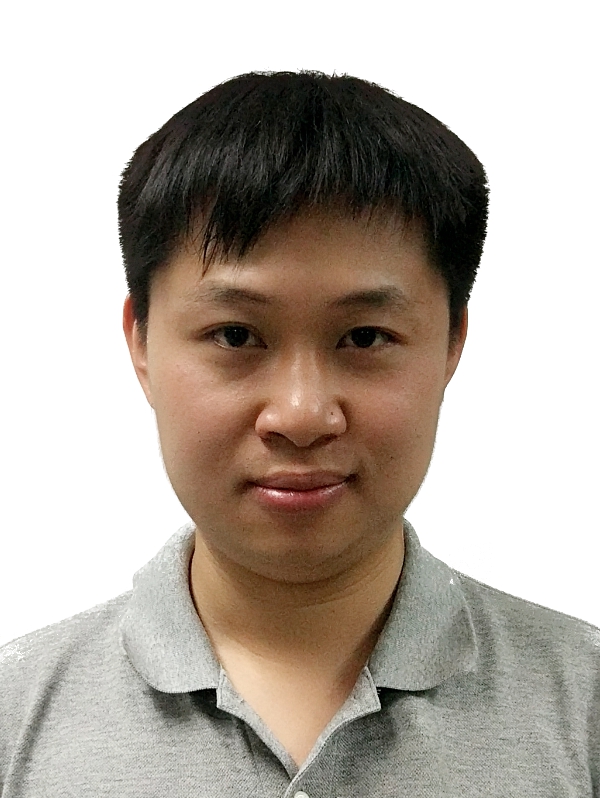}}]{Qinyong Wang} is currently a Ph.D. student at The University of Queensland. He obtained Master Degree from Chinese Academy of Sciences in 2017 and Bachelor Degree  from Central South University in 2014. His research interests include recommender systems, knowledge graphs, network embedding and social-media analysis. His work has been published in some of the most prestigious venues such as KDD, WWW, ICDE, IJCAI and SIGIR. He received the best paper awards at ICDE 2019 and ADC 2018. 
\end{IEEEbiography}
\vskip -2\baselineskip plus -1fil
\begin{IEEEbiography}[{\includegraphics[width=1in,height=1.15in,clip,keepaspectratio]{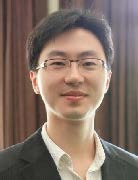}}]{Kai Zheng}  is a Professor of Computer Science with University of Electronic Science and Technology of China. He received his PhD degree in Computer Science from The University of Queensland in 2012. He has been working in the area of spatial-temporal databases, uncertain databases, social-media analysis, in memory computing and block chain technologies. He has published over 100 papers in prestigious journals and conferences in data management field such as SIGMOD, ICDE, VLDB Journal, ACM Transactions and IEEE Transactions. He is a member of IEEE. 
\end{IEEEbiography}
\vskip -2\baselineskip plus -1fil
\begin{IEEEbiography}[{\includegraphics[width=1in,height=1.15in,clip,keepaspectratio]{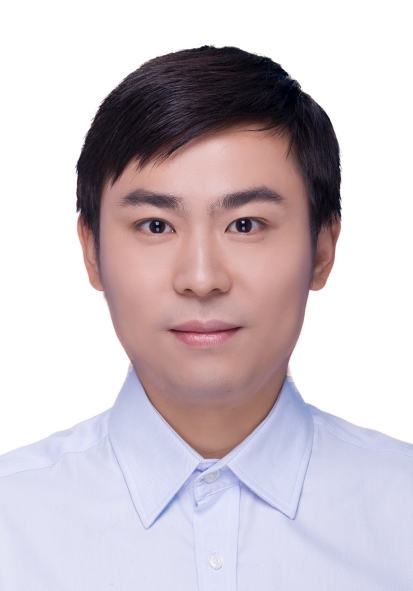}}]{Zhixu Li} is an associate professor in the Department of Computer Science and Technology at Soochow University, China. He used to work as a research fellow at KAUST. He received his Ph.D. degree from the University of Queensland in 2013, and his B.S. and M.S. degree from Renmin University of China in 2006 and 2009 respectively. His research interests are Knowledge Graph, Question Answering, Data Quality and various Big Data Applications. 
\end{IEEEbiography}
\vskip -2\baselineskip plus -1fil
\begin{IEEEbiography}[{\includegraphics[width=1in,height=1.15in,clip,keepaspectratio]{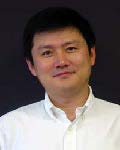}}]{Xiaofang Zhou} received the BSc and MSc degrees in computer science from Nanjing University, China, in 1984 and 1987, respectively, and the PhD degree in computer science from the University of Queensland, Australia, in 1994. He is a professor of computer science at the University of Queensland and adjunct professor in the School of Computer Science and Technology, Soochow University, China. His research interests include spatial and multimedia databases, high performance query processing, web information systems, data mining, bioinformatics, and e-research. He is a fellow of the IEEE.
\end{IEEEbiography}

\end{document}